# Accurate De Novo Prediction of Protein Contact Map by Ultra-Deep Learning Model

Sheng Wang, Siqi Sun, Zhen Li, Renyu Zhang and Jinbo Xu*

Toyota Technological Institute at Chicago

jinboxu@gmail.com

*: corresponding author

The first two authors contribute equally

## Abstract

**Motivation:** Protein contacts contain key information for the understanding of protein structure and function and thus, contact prediction from sequence is an important problem. Recently exciting progress has been made on this problem, but the predicted contacts for proteins without many sequence homologs is still of low quality and not extremely useful for de novo structure prediction.

**Method:** This paper presents a new deep learning method that predicts contacts by integrating both evolutionary coupling (EC) and sequence conservation information through an ultra-deep neural network formed by two deep residual neural networks. The first residual network conducts a series of 1-dimensional convolutional transformation of sequential features; the second residual network conducts a series of 2-dimensional convolutional transformation of pairwise information including output of the first residual network, EC information and pairwise potential. By using very deep residual networks, we can accurately model contact occurring patterns and complex sequence-structure relationship and thus, obtain high-quality contact prediction regardless of how many sequence homologs are available for proteins in question.

**Results:** Our method greatly outperforms existing methods and leads to much more accurate contact-assisted folding. Tested on 105 CASP11 targets, 76 past CAMEO hard targets, and 398 membrane proteins, the average top L long-range prediction accuracy obtained our method, one representative EC method CCMpred and the CASP11 winner MetaPSICOV is 0.47, 0.21 and 0.30, respectively; the average top L/10 long-range accuracy of our method, CCMpred and MetaPSICOV is 0.77, 0.47 and 0.59, respectively. Ab initio folding using our predicted contacts as restraints but without any force fields can yield correct folds (i.e., TMscore>0.6) for 203 of the 579 test proteins, while that using MetaPSICOV- and CCMpred-predicted contacts can do so for only 79 and 62 of them, respectively. Our contact-assisted models also have much better quality than template-based models especially for membrane proteins. The 3D models built from our contact prediction have TMscore>0.5 for 208 of the 398 membrane proteins, while those from homology modeling have TMscore>0.5 for only 10 of them. Further, even if trained by only non-membrane proteins, our deep learning method works very well on membrane protein contact prediction. In the recent blind CAMEO benchmark, our fully-automated web server implementing this method successfully folded 5 targets with a new fold and only 0.3L-2.3L effective sequence homologs, including one β protein of 182 residues, one α+β protein of 125 residues, one α protein of 140 residues, one α protein of 217 residues and one α/β of 260 residues.

**Availability:** http://raptorx.uchicago.edu/ContactMap/

## Author Summary

Protein contact prediction and contact-assisted folding has made good progress due to direct evolutionary coupling analysis (DCA). However, DCA is effective on only some proteins with a very large number of sequence homologs. To further improve contact prediction, we borrow ideas from deep learning, which has recently revolutionized object recognition, speech recognition and the GO game. Our deep learning method can model complex sequence-structure relationship and high-order correlation (i.e., contact occurring patterns) and thus, improve contact prediction accuracy greatly. Our test results show that our method greatly outperforms the state-of-the-art methods regardless how many sequence homologs are available for a protein in question. Ab initio folding guided by our predicted contacts may fold many more test proteins than the other contact predictors. Our contact-assisted 3D models also have much better quality than homology models built from the training proteins, especially for membrane proteins. One interesting finding is that even trained with only soluble proteins, our method performs very well on membrane proteins. Recent blind test in CAMEO confirms that our method can fold large proteins with a new fold and only a small number of sequence homologs.

## Introduction

De novo protein structure prediction from sequence alone is one of most challenging problems in computational biology. Recent progress has indicated that some correctly-predicted long-range contacts may allow accurate topology-level structure modeling (1) and that direct evolutionary coupling analysis (DCA) of multiple sequence alignment (MSA) may reveal some long-range native contacts for proteins and protein-protein interactions with a large number of sequence homologs (2, 3). Therefore, contact prediction and contact-assisted protein folding has recently gained much attention in the community. However, for many proteins especially those without many sequence homologs, the predicted contacts by the state-of-the-art predictors such as CCMpred (4), PSICOV (5), Evfold (6), plmDCA(7), Gremlin(8), MetaPSICOV (9) and CoinDCA (10) are still of low quality and insufficient for accurate contact-assisted protein folding (11,12). This motivates us to develop a better contact prediction method, especially for proteins without a large number of sequence homologs. In this paper we define that two residues form a contact if they are spatially proximal in the native structure, i.e., the Euclidean distance of their $C_\beta$ atoms less than 8Å (13).

Existing contact prediction methods roughly belong to two categories: evolutionary coupling analysis

(ECA) and supervised machine learning. ECA predicts contacts by identifying co-evolved residues in a protein, such as EVfold (6), PSICOV (5), CCMpred (4), Gremlin (8), plmDCA and others (14-16). However, DCA usually needs a large number of sequence homologs to be effective (10, 17). Supervised machine learning predicts contacts from a variety of information, e.g., SVMSEQ (18), CMAPpro (13), PconsC2 (17), MetaPSICOV (9), PhyCMAP (19) and CoinDCA-NN (10). Meanwhile, PconsC2 uses a 5-layer supervised learning architecture (17); CoinDCA-NN and MetaPSICOV employ a 2-layer neural network (9). CMAPpro uses a neural network with more layers, but its performance saturates at about 10 layers. Some supervised methods such as MetaPSICOV and CoinDCA-NN outperform ECA on proteins without many sequence homologs, but their performance is still limited by their shallow architectures.

To further improve supervised learning methods for contact prediction, we borrow ideas from very recent breakthrough in computer vision. In particular, we have greatly improved contact prediction by developing a brand-new deep learning model called residual neural network (20) for contact prediction. Deep learning is a powerful machine learning technique that has revolutionized image classification (21, 22) and speech recognition (23). In 2015, ultra-deep residual neural networks (24) demonstrated superior performance in several computer vision challenges (similar to CASP) such as image classification and object recognition (25). If we treat a protein contact map as an image, then protein contact prediction is kind of similar to (but not exactly same as) pixel-level image labeling, so some techniques effective for image labeling may also work for contact prediction. However, there are some important differences between image labeling and contact prediction. First, in computer vision community, image-level labeling (i.e., classification of a single image) has been extensively studied, but there are much fewer studies on pixel-level image labeling (i.e., classification of an individual pixel). Second, in many image classification scenarios, image size is resized to a fixed value, but we cannot resize a contact map since we need to do prediction for every residue pair (equivalent to an image pixel). Third, contact prediction has much more complex input features (including both sequential and pairwise features) than image labeling. Fourth, the ratio of contacts in a protein is very small (<2%). That is, the number of positive and negative labels in contact prediction is extremely unbalanced.

In this paper we present a very deep residual neural network for contact prediction. Such a network can capture very complex sequence-contact relationship and high-order contact correlation. We train this deep neural network using a subset of proteins with solved structures and then test its performance on public data including the CASP (26, 27) and CAMEO (28) targets as well as many membrane proteins. Our experimental results show that our method yields much better accuracy than existing methods and also result in much more accurate contact-assisted folding. The deep learning method described here will also be useful for the prediction of protein-protein and protein-RNA interfacial contacts.

# Results

## Deep learning model for contact prediction

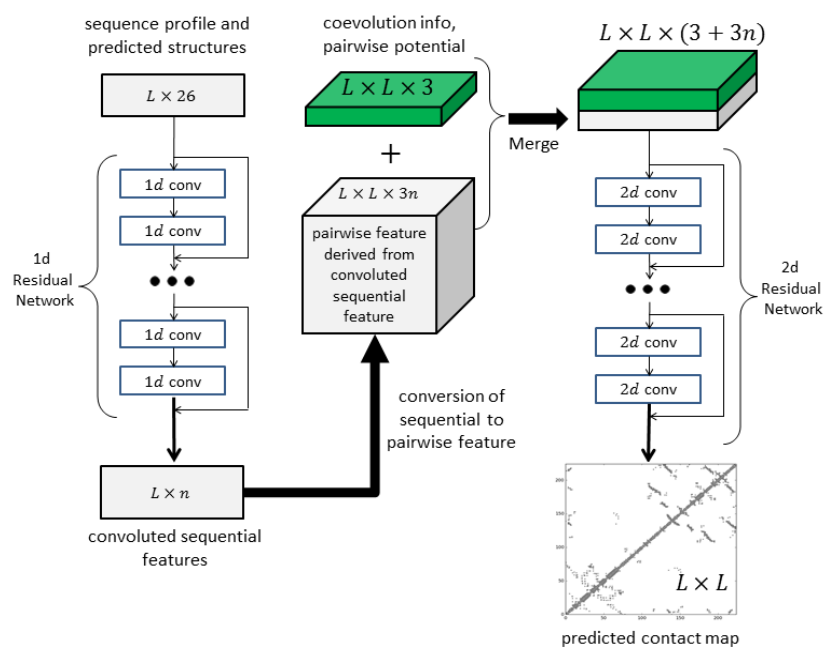

**Figure 1.** Illustration of our deep learning model for contact prediction. Meanwhile, L is the sequence length of one protein under prediction.

Fig. 1 illustrates our deep neural network model for contact prediction (29). Different from previous supervised learning approaches(9, 13) for contact prediction that employ only a small number of hidden layers (i.e., a shallow architecture), our deep neural network employs dozens of hidden layers. By using a very deep architecture, our model can automatically learn the complex relationship between sequence information and contacts and also model the interdependency among contacts and thus, improve contact prediction (17). Our model consists of two major modules, each being a residual neural network. The first module conducts a series of 1-dimensional (1D) convolutional transformations of sequential features (sequence profile, predicted secondary structure and solvent accessibility). The output of this 1D convolutional network is converted to a 2-dimensional (2D) matrix by an operation similar to outer product and then fed into the $2^{nd}$ module together with pairwise features (i.e., co-evolution information, pairwise contact and distance potential). The $2^{nd}$ module is a 2D residual network that conducts a series of 2D convolutional transformations of its input. Finally, the output of the 2D convolutional network is fed into a logistic regression, which predicts the probability of any two residues form a contact. In addition, each convolutional layer is also preceded by a simple nonlinear transformation called rectified linear unit (30). Mathematically, the output of 1D residual network is just a 2D matrix with dimension L×m where m is the number of new features (or hidden neurons) generated by the last convolutional layer of the network. Biologically, this 1D residual network learns the sequential context of a residue. By stacking multiple convolution layers, the network can learn information in a very large sequential context. The output of a 2D convolutional layer has dimension L×L×n where n is the number of new features (or hidden neurons) generated by this layer for one residue pair. The 2D residual network mainly learns contact occurring patterns or high-order residue correlation (i.e., 2D context of a residue pair). The number of hidden neurons may vary at each layer.

Our test data includes the 150 Pfam families described in (5), 105 CASP11 test proteins (31), 398 membrane proteins (Supplementary Table 1) and 76 CAMEO hard targets released from 10/17/2015 to 04/09/2016 (Supplementary Table 2). The tested methods include PSICOV (5), Evfold (6), CCMpred (4), plmDCA(7), Gremlin(8), and MetaPSICOV (9). The former 5 methods employs pure DCA while MetaPSICOV (9) is a supervised learning method that performed the best in CASP11 (31). All the programs are run with parameters set according to their respective papers. We cannot evaluate PconsC2 (17) since we failed to obtain any results from its web server. PconsC2 did not outperform MetaPSICOV in CASP11 (31), so it may suffice to just compare our method with MetaPSICOV.

## Overall Performance

We evaluate the accuracy of the top $L/k$ ($k$=10, 5, 2, 1) predicted contacts where L is protein sequence length (10). We define that a contact is short-, medium- and long-range when the sequence distance of the two residues in a contact falls into [6, 11], [12, 23], and ≥24, respectively. The prediction accuracy is defined as the percentage of native contacts among the top $L/k$ predicted contacts. When there are no $L/k$ native (short- or medium-range) contacts, we replace the denominator by $L/k$ in calculating accuracy. This may make the short- and medium-range accuracy look small although it is easier to predict short- and medium-range contacts than long-range ones.

**Table 1.** Contact prediction accuracy on the 150 Pfam families.

| Method | Short | | | | Medium | | | | Long | | | |
|---|---|---|---|---|---|---|---|---|---|---|---|---|
| | L/10 | L/5 | L/2 | L | L/10 | L/5 | L/2 | L | L/10 | L/5 | L/2 | L |
| EVfold | 0.50 | 0.40 | 0.26 | 0.17 | 0.64 | 0.52 | 0.34 | 0.22 | 0.74 | 0.68 | 0.53 | 0.39 |
| PSICOV | 0.58 | 0.43 | 0.26 | 0.17 | 0.65 | 0.51 | 0.32 | 0.20 | 0.77 | 0.70 | 0.52 | 0.37 |
| CCMpred | 0.65 | 0.50 | 0.29 | 0.19 | 0.73 | 0.60 | 0.37 | 0.23 | 0.82 | 0.76 | 0.62 | 0.45 |
| plmDCA | 0.66 | 0.50 | 0.29 | 0.19 | 0.72 | 0.60 | 0.36 | 0.22 | 0.81 | 0.76 | 0.61 | 0.44 |
| Gremlin | 0.66 | 0.51 | 0.30 | 0.19 | 0.74 | 0.60 | 0.37 | 0.23 | 0.82 | 0.76 | 0.63 | 0.46 |
| MetaPSICOV | 0.82 | 0.70 | 0.45 | 0.27 | 0.83 | 0.73 | 0.52 | 0.33 | 0.92 | 0.87 | 0.74 | 0.58 |
| Our method | 0.93 | 0.81 | 0.51 | 0.30 | 0.93 | 0.86 | 0.62 | 0.38 | 0.98 | 0.96 | 0.89 | 0.74 |

**Table 2.** Contact prediction accuracy on 105 CASP11 test proteins.

| Method | Short | | | | Medium | | | | Long | | | |
|---|---|---|---|---|---|---|---|---|---|---|---|---|
| | L/10 | L/5 | L/2 | L | L/10 | L/5 | L/2 | L | L/10 | L/5 | L/2 | L |
| EVfold | 0.25 | 0.21 | 0.15 | 0.12 | 0.33 | 0.27 | 0.19 | 0.13 | 0.37 | 0.33 | 0.25 | 0.19 |
| PSICOV | 0.29 | 0.23 | 0.15 | 0.12 | 0.34 | 0.27 | 0.18 | 0.13 | 0.38 | 0.33 | 0.25 | 0.19 |

| Method | | | | | | | | | | | | |
|---|---|---|---|---|---|---|---|---|---|---|---|---|
| CCMpred | 0.35 | 0.28 | 0.17 | 0.12 | 0.40 | 0.32 | 0.21 | 0.14 | 0.43 | 0.39 | 0.31 | 0.23 |
| plmDCA | 0.32 | 0.26 | 0.17 | 0.12 | 0.39 | 0.31 | 0.21 | 0.14 | 0.42 | 0.38 | 0.30 | 0.23 |
| Gremlin | 0.35 | 0.27 | 0.17 | 0.12 | 0.40 | 0.31 | 0.21 | 0.14 | 0.44 | 0.40 | 0.31 | 0.23 |
| MetaPSICOV | 0.69 | 0.58 | 0.39 | 0.25 | 0.69 | 0.59 | 0.42 | 0.28 | 0.60 | 0.54 | 0.45 | 0.35 |
| Our method | 0.82 | 0.70 | 0.46 | 0.28 | 0.85 | 0.76 | 0.55 | 0.35 | 0.81 | 0.77 | 0.68 | 0.55 |

**Table 3.** Contact prediction accuracy on 76 past CAMEO hard targets.

| Method | Short | | | | Medium | | | | Long | | | |
|---|---|---|---|---|---|---|---|---|---|---|---|---|
| | L/10 | L/5 | L/2 | L | L/10 | L/5 | L/2 | L | L/10 | L/5 | L/2 | L |
| EVfold | 0.17 | 0.13 | 0.11 | 0.09 | 0.23 | 0.19 | 0.13 | 0.10 | 0.25 | 0.22 | 0.17 | 0.13 |
| PSICOV | 0.20 | 0.15 | 0.11 | 0.08 | 0.24 | 0.19 | 0.13 | 0.09 | 0.25 | 0.23 | 0.18 | 0.13 |
| CCMpred | 0.22 | 0.16 | 0.11 | 0.09 | 0.27 | 0.22 | 0.14 | 0.10 | 0.30 | 0.26 | 0.20 | 0.15 |
| plmDCA | 0.23 | 0.18 | 0.12 | 0.09 | 0.27 | 0.22 | 0.14 | 0.10 | 030 | 0.26 | 0.20 | 0.15 |
| Gremlin | 0.21 | 0.17 | 0.11 | 0.08 | 0.27 | 0.22 | 0.14 | 0.10 | 0.31 | 0.26 | 0.20 | 0.15 |
| MetaPSICOV | 0.56 | 0.47 | 0.31 | 0.20 | 0.53 | 0.45 | 0.32 | 0.22 | 0.47 | 0.42 | 0.33 | 0.25 |
| Our method | 0.67 | 0.57 | 0.37 | 0.23 | 0.69 | 0.61 | 0.42 | 0.28 | 0.69 | 0.65 | 0.55 | 0.42 |

**Table 4.** Contact prediction accuracy on 398 membrane proteins.

| Method | Short | | | | Medium | | | | Long | | | |
|---|---|---|---|---|---|---|---|---|---|---|---|---|
| | L/10 | L/5 | L/2 | L | L/10 | L/5 | L/2 | L | L/10 | L/5 | L/2 | L |
| EVfold | 0.16 | 0.13 | 0.09 | 0.07 | 0.28 | 0.22 | 0.13 | 0.09 | 0.44 | 0.37 | 0.26 | 0.18 |
| PSICOV | 0.22 | 0.16 | 0.10 | 0.07 | 0.29 | 0.21 | 0.13 | 0.09 | 0.42 | 0.34 | 0.23 | 0.16 |
| CCMpred | 0.27 | 0.19 | 0.11 | 0.08 | 0.36 | 0.26 | 0.15 | 0.10 | 0.52 | 0.45 | 0.31 | 0.21 |
| plmDCA | 0.26 | 0.18 | 0.11 | 0.08 | 0.35 | 0.25 | 0.14 | 0.09 | 0.51 | 0.42 | 0.29 | 0.20 |
| Gremlin | 0.27 | 0.19 | 0.11 | 0.07 | 0.37 | 0.26 | 0.15 | 0.10 | 0.52 | 0.45 | 0.32 | 0.21 |
| MetaPSICOV | 0.45 | 0.35 | 0.22 | 0.14 | 0.49 | 0.40 | 0.27 | 0.18 | 0.61 | 0.55 | 0.42 | 0.30 |
| Our method | 0.60 | 0.46 | 0.27 | 0.16 | 0.66 | 0.53 | 0.33 | 0.22 | 0.78 | 0.73 | 0.62 | 0.47 |

As shown in Tables 1-4, our method outperforms all tested DCA methods and MetaPSICOV by a very large margin on the 4 test sets regardless of how many top predicted contacts are evaluated and no matter whether the contacts are short-, medium- or long-range. These results also show that two supervised learning methods greatly outperform the pure DCA methods and the three pseudo-likelihood DCA methods plmDCA, Gremlin and CCMpred perform similarly, but outperform PSICOV (Gaussian model) and Evfold (maximum-entropy method). The advantage of our method is the smallest on the 150 Pfam families because many of them have a pretty large number of sequence homologs. In terms of top L long-range contact accuracy on the CASP11 set, our method exceeds CCMpred and MetaPSICOV by 0.32 and 0.20, respectively. On the 76 CAMEO hard targets, our method exceeds CCMpred and MetaPSICOV by 0.27 and 0.17, respectively. On the 398 membrane protein set, our method exceeds CCMpred and MetaPSICOV by 0.26 and 0.17, respectively. Our method uses a subset of protein features used by MetaPSICOV, but performs much better than MetaPSICOV due to our deep architecture and that we predict contacts of a protein simultaneously. Since the Pfam set is relatively easy, we will not analyze it any more in the following sections.

## Prediction accuracy with respect to the number of sequence homologs

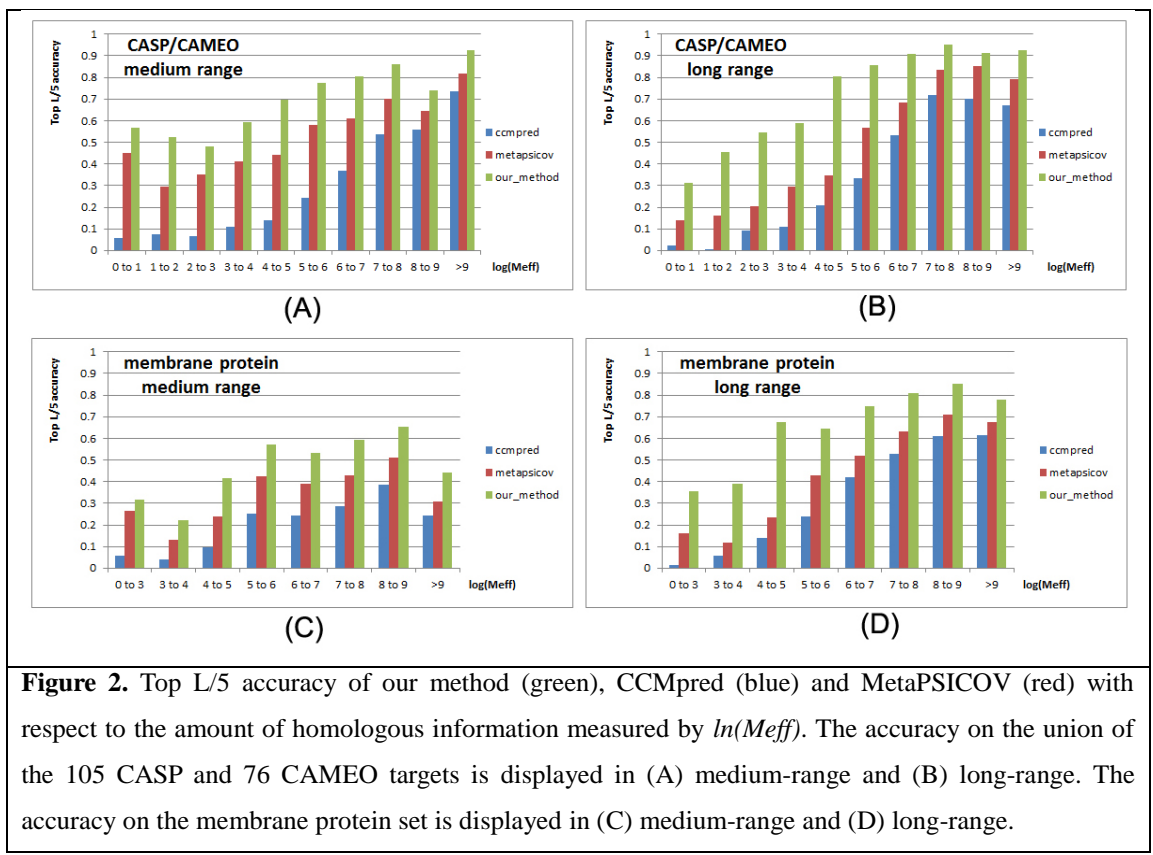

**Figure 2.** Top L/5 accuracy of our method (green), CCMpred (blue) and MetaPSICOV (red) with respect to the amount of homologous information measured by *ln(Meff)*. The accuracy on the union of the 105 CASP and 76 CAMEO targets is displayed in (A) medium-range and (B) long-range. The accuracy on the membrane protein set is displayed in (C) medium-range and (D) long-range.

To examine the performance of our method with respect to the amount of homologous information available for a protein under prediction, we measure the effective number of sequence homologs in multiple sequence alignment (MSA) by *Meff* (19), which can be roughly interpreted as the number of non-redundant sequence homologs when 70% sequence identity is used as cutoff to remove redundancy (see Method for its formula). A protein with a smaller *Meff* has less homologous information. We divide all the test proteins into 10 bins according to *ln(Meff)* and then calculate the average accuracy of proteins in each bin. We merge the first 3 bins for the membrane protein set since they have a small number of proteins.

Fig. 2 shows that the top L/5 contact prediction accuracy increases with respect to *Meff*, i.e., the number of effective sequence homologs, and that our method outperforms both MetaPSICOV and CCMpred regardless of *Meff*. Our long-range prediction accuracy is even better when *ln(Meff)≤7* (equivalently *Meff<1100*), i.e., when the protein under prediction does not have a very large number of non-redundant sequence homologs. Our method has a large advantage over the other methods even when *Meff* is very big (>8000). This indicates that our method indeed benefits from some extra information such as inter-contact correlation or high-order residue correlation, which is orthogonal to pairwise co-evolution information.

## Contact-assisted protein folding

One of the important goals of contact prediction is to perform contact-assisted protein folding (11). To test if our contact prediction can lead to better 3D structure modeling than the others, we build structure models for all the test proteins using the top predicted contacts as restraints of ab initio folding. For each test protein, we feed the top predicted contacts as restraints into the CNS suite (32) to generate 3D models. We measure the quality of a 3D model by a superposition-dependent score TMscore (33), which ranges from 0 to 1, with 0 indicating the worst and 1 the best, respectively. We also measure the quality of a 3D model by a superposition-independent score lDDT, which ranges from 0 to 100, with 0 indicating the worst and 100 the best, respectively.

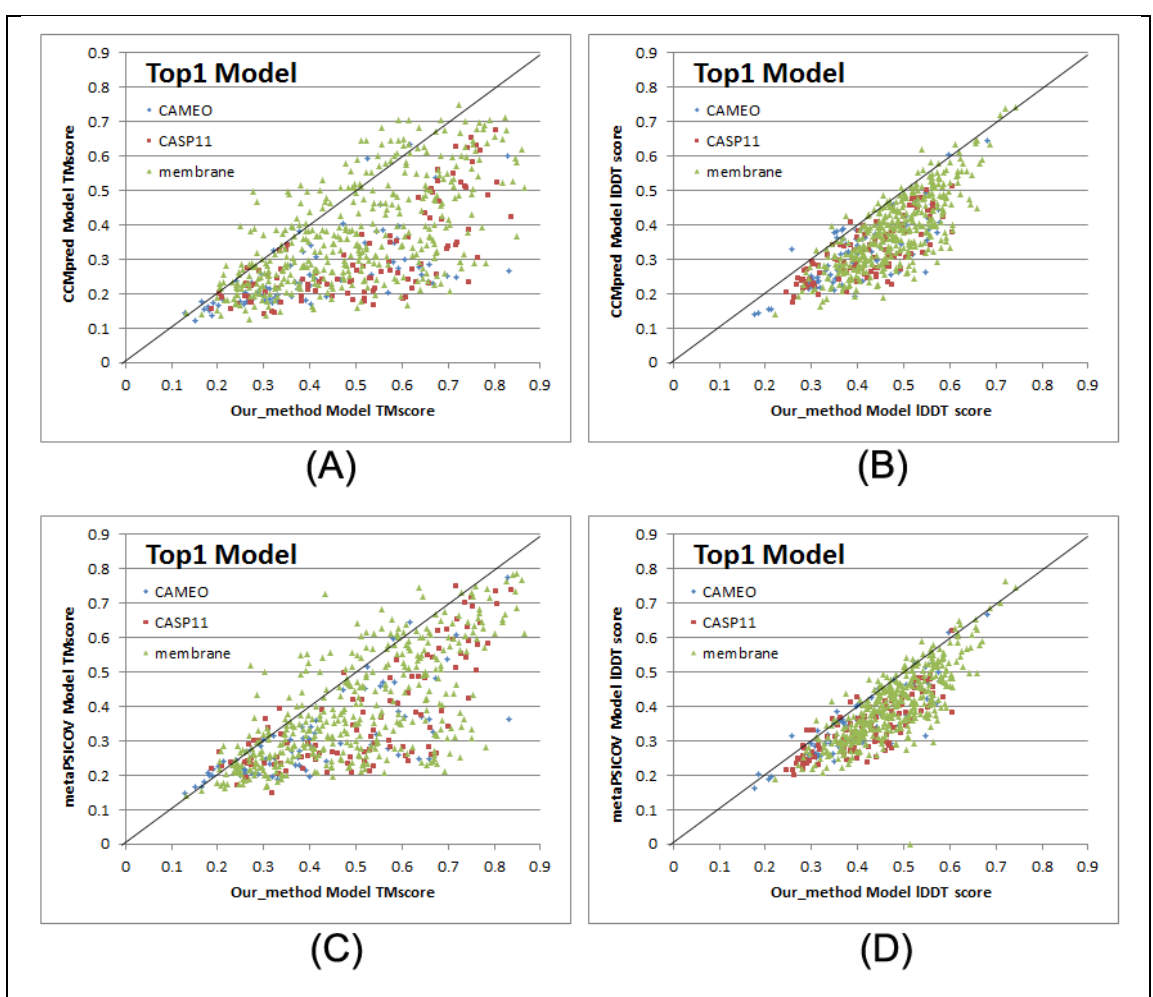

**Figure 3.** Quality comparison of top 1 contact-assisted models generated by our method, CCMpred and MetaPSICOV on the 105 CASP11 targets (red square), 76 CAMEO targets (blue diamond) and 398 membrane protein targets (green triangle), respectively. **(A)** and **(B)**: comparison between our method (X-axis) and CCMpred (Y-axis) in terms of TMscore and lDDT, respectively. **(C)** and **(D)**: comparison between our method (X-axis) and MetaPSICOV (Y-axis) in terms of TMscore and lDDT, respectively.

> lDDT is scaled to between 0 and 1.

Fig. 3 shows that our predicted contacts can generate much better 3D models than CCMpred and MetaPSICOV. On average, our 3D models are better than MetaPSICOV and CCMpred by ~0.12 TMscore unit and ~0.15 unit, respectively. When the top 1 models are evaluated, the average TMscore obtained by CCMpred, MetaPSICOV, and our method is 0.333, 0.377, and 0.518, respectively on the CASP dataset. The average lDDT of CCMpred, MetaPSICOV and our method is 31.7, 34.1 and 41.8, respectively. On the 76 CAMEO targets, the average TMsore of CCMpred, MetaPSICOV and our method is 0.256, 0.305 and 0.407, respectively. The average lDDT of CCMpred, MetaPSICOV and our method is 31.8, 35.4 and 40.2, respectively. On the membrane protein set, the average TMscore of CCMpred, MetaPSICOV and our method is 0.354, 0.387, and 0.493, respectively. The average lDDT of CCMpred, MetaPSICOV and our method is 38.1, 40.5 and 47.8, respectively. Same trend is observed when the best of top 5 models are evaluated (Supplementary Figure 1). On the CASP set, the average TMscore of the models generated by CCMpred, MetaPSICOV, and our method is 0.352, 0.399, and 0.543, respectively. The average lDDT of CCMpred, MetaPSICOV and our method is 32.3, 34.9 and 42.4, respectively. On the 76 CAMEO proteins, the average TMscore of CCMpred, MetaPSICOV, and our method is 0.271, 0.334, and 0.431, respectively. The average lDDT of CCMpred, MetaPSICOV and our method is 32.4, 36.1 and 40.9, respectively. On the membrane protein set, the average TMscore of CCMpred, MetaPSICOV, and our method is 0.385, 0.417, and 0.516, respectively. The average lDDT of CCMpred, MetaPSICOV and our method is 38.9, 41.2 and 48.5, respectively. In particular, when the best of top 5 models are considered, our predicted contacts can result in correct folds (i.e., TMscore>0.6) for 203 of the 579 test proteins, while MetaPSICOV- and CCMpred-predicted contacts can do so for only 79 and 62 of them, respectively.

Our method also generates much better contact-assisted models for the test proteins without many non-redundant sequence homologs. When the 219 of 579 test proteins with $M_{eff}$≤500 are evaluated, the average TMscore of the top 1 models generated by our predicted contacts for the CASP11, CAMEO and membrane sets is 0.426, 0.365, and 0.397, respectively. By contrast, the average TMscore of the top 1 models generated by CCMpred-predicted contacts for the CASP11, CAMEO and membrane sets is 0.236, 0.214, and 0.241, respectively. The average TMscore of the top 1 models generated by MetaPSICOV-predicted contacts for the CASP11, CAMEO and membrane sets is 0.292, 0.272, and 0.274, respectively.

## Contact-assisted models vs. template-based models

To compare the quality of our contact-assisted models and template-based models (TBMs), we built TBMs for all the test proteins using our training proteins as candidate templates. To generate TBMs for a test protein, we first run HHblits (with the UniProt20_2016 library) to generate an HMM file for the test protein, then run HHsearch with this HMM file to search for the best templates among the 6767 training proteins, and finally run MODELLER to build a TBM from each of the top 5 templates. Fig. 4 shows the head-to-head comparison between our top 1 contact-assisted models and the top 1 TBMs on these three test sets in terms of both TMscore and lDDT. The average lDDT of our top 1 contact-assisted models is 45.7 while that of top 1 TBMs is only 20.7. When only the first models are evaluated, our contact-assisted models for the 76 CAMEO test proteins have an average TMscore 0.407 while the TBMs have an average TMscore 0.317. On the 105 CASP11 test proteins, the average TMscore of our contact-assisted models is 0.518 while that of the TBMs is only 0.393. On the 398 membrane proteins, the average TMscore of our contact-assisted models is 0.493 while that of the TBMs is only 0.149. Same trend is observed when top 5 models are compared (see Supplementary Figure 2). The average lDDT of our top 5 contact-assisted models is 46.4 while that of top 5 TBMs is only 24.0. On the 76 CAMEO test proteins, the average TMscore of our contact-assisted models is 0.431 while that of the TBMs is only 0.366. On the 105 CASP11 test proteins, the average TMscore of our contact-assisted models is 0.543 while that of the TBMs is only 0.441. On the 398 membrane proteins, the average TMscore of our contact-assisted models is 0.516 while that of the TBMs is only 0.187. The low quality of TBMs further confirms that there is little redundancy between our training and test proteins (especially membrane proteins). This also indicates that our deep model does not predict contacts by simply copying from training proteins. That is, our method can predict contacts for a protein with a new fold.

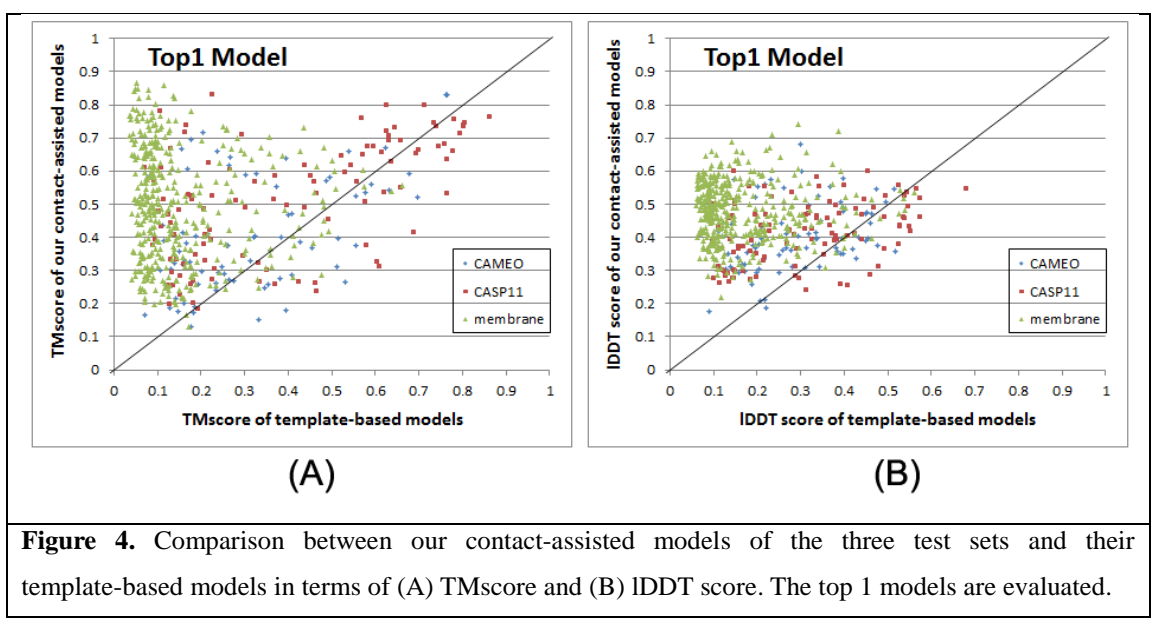

**Figure 4.** Comparison between our contact-assisted models of the three test sets and their template-based models in terms of (A) TMscore and (B) lDDT score. The top 1 models are evaluated.

Further, when the best of top 5 models are considered for all the methods, our contact-assisted models have TMscore>0.5 for 24 of the 76 CAMEO targets while TBMs have TMscore>0.5 for only 18 of them. Our contact-assisted models have TMscore >0.5 for 67 of the 105 CASP11 targets while TBMs have TMscore>0.5 for only 44 of them. Our contact-assisted models have TMscore>0.5 for 208 of the 398 membrane proteins while TBMs have TMscore >0.5 for only 10 of them. Our contact-assisted models for membrane proteins are much better than their TBMs because there is little similarity between the 6767 training proteins and the 398 test membrane proteins. When the 219 test proteins with ≤500 non-redundant sequence homologs are evaluated, the average TMscore of the TBMs is 0.254 while that of our contact-assisted models is 0.421. Among these 219 proteins, our contact-assisted models have TMscore>0.5 for 72 of them while TBMs have TMscore>0.5 for only 17 of them.

The above results imply that 1) when a query protein has no close templates, our contact-assisted modeling may work better than template-based modeling; 2) contact-assisted modeling shall be particularly useful for membrane proteins; and 3) our deep learning model does not predict contacts by simply copying contacts from the training proteins since our predicted contacts may result in much better 3D models than homology modeling.

## Blind test in CAMEO

We have implemented our algorithm as a fully-automated contact prediction web server (http://raptorx.uchicago.edu/ContactMap/) and in September 2016 started to blindly test it through the weekly live benchmark CAMEO (http://www.cameo3d.org/). CAMEO is operated by the Schwede group, with whom we have never collaborated. CAMEO can be interpreted as a fully-automated CASP, but has a smaller number (>20) of participating servers since many CASP-participating servers are not fully automated and thus, cannot handle the large number of test targets used by CAMEO. Nevertheless, the CAMEO participants include some well-known servers such as Robetta(34), Phyre(35), RaptorX(36), Swiss-Model(37) and HHpred(38). Meanwhile Robetta employs both ab initio folding and template-based modeling while the latter four employ mainly template-based modeling. Every weekend CAMEO sends test sequences to participating servers for prediction and then evaluates 3D models collected from servers. The test proteins used by CAMEO have no publicly available native structures until CAMEO finishes collecting models from participating servers.

During the past 2 months (9/3/2016 to 10/31/2016), CAMEO in total released 41 hard targets (Supplementary Table 3). Although classified as hard by CAMEO, some of them may have distantly-related templates. Table 5 lists the contact prediction accuracy of our server in the blind CAMEO test as compared to the other methods. Again, our method outperforms the others by a very large margin no matter how many contacts are evaluated. The CAMEO evaluation of our contact-assisted 3D models is available at the CAMEO web site. You will need to register CAMEO in order to see all the detailed results of our contact server (ID: server60). Although our server currently build 3D models using only top predicted contacts without any force fields and fragment assembly procedures, our server predicts 3D models with TMscore>0.5 for 28 of the 41 targets and TMscore>0.6 for 16 of them. The average TMscore of the best of top 5 models built from the contacts predicted by our server, CCMpred and MetaPSICOV is 0.535, 0.316 and 0.392, respectively. See Fig. 5 for the detailed comparison of the 3D models generated by our server, CCMpred and MetaPSICOV. Our server has also successfully folded 4 targets with a new fold plus one released in November 2016 (5flgB). See Table 6 for a summary of our prediction results of these targets and the below subsections for a detailed analysis. Among these targets, 5f5pH is particularly interesting since it has a sequence homolog in PDB but adopting a different conformation. That is, any template-based techniques cannot obtain a good prediction for this target.

Table 5. Contact prediction accuracy on 41 recent CAMEO hard targets.

| Method | Short | | | | Medium | | | | Long | | | |
|---|---|---|---|---|---|---|---|---|---|---|---|---|
| | L/10 | L/5 | L/2 | L | L/10 | L/5 | L/2 | L | L/10 | L/5 | L/2 | L |
| EVfold | 0.20 | 0.15 | 0.11 | 0.08 | 0.25 | 0.19 | 0.12 | 0.09 | 0.33 | 0.29 | 0.21 | 0.15 |
| PSICOV | 0.21 | 0.16 | 0.11 | 0.08 | 0.26 | 0.20 | 0.11 | 0.08 | 0.33 | 0.30 | 0.21 | 0.15 |
| **plmDCA** | 0.26 | 0.19 | 0.12 | 0.09 | 0.28 | 0.23 | 0.13 | 0.09 | 0.38 | 0.33 | 0.24 | 0.17 |
| **Gremlin** | 0.25 | 0.18 | 0.12 | 0.08 | 0.29 | 0.22 | 0.13 | 0.09 | 0.37 | 0.34 | 0.25 | 0.17 |
| CCMpred | 0.24 | 0.18 | 0.12 | 0.08 | 0.29 | 0.22 | 0.13 | 0.09 | 0.37 | 0.34 | 0.24 | 0.17 |
| MetaPSICOV | 0.53 | 0.43 | 0.27 | 0.17 | 0.51 | 0.42 | 0.28 | 0.19 | 0.60 | 0.54 | 0.40 | 0.30 |
| Our server | 0.67 | 0.52 | 0.32 | 0.20 | 0.68 | 0.58 | 0.38 | 0.24 | 0.82 | 0.75 | 0.62 | 0.46 |

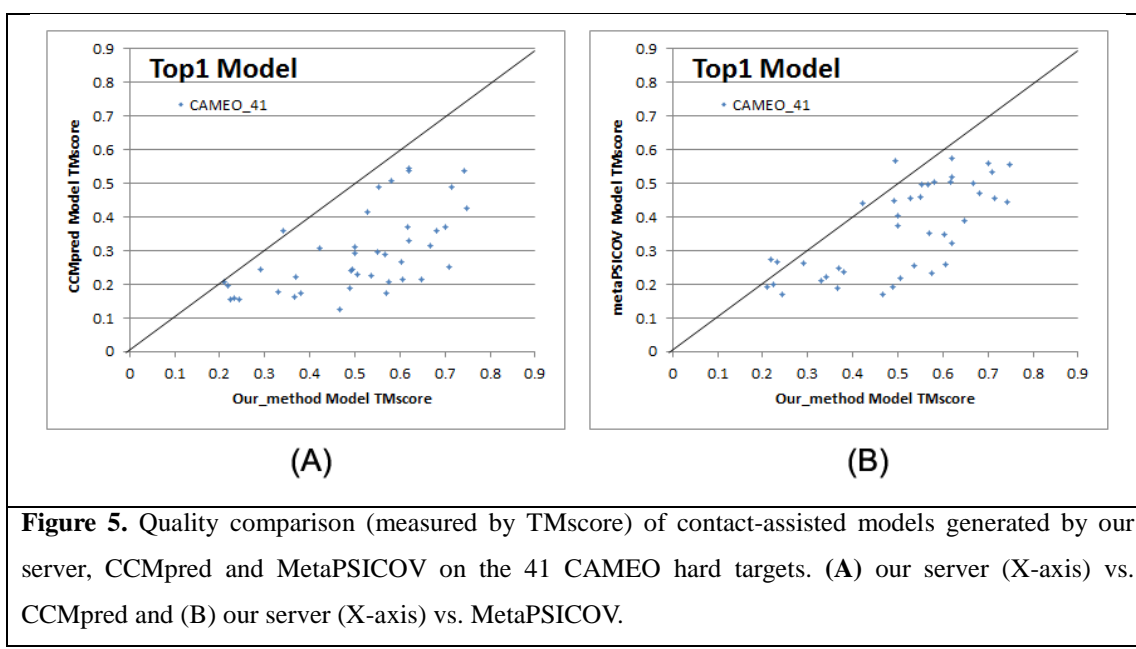

**Figure 5.** Quality comparison (measured by TMscore) of contact-assisted models generated by our server, CCMpred and MetaPSICOV on the 41 CAMEO hard targets. **(A)** our server (X-axis) vs. CCMpred and (B) our server (X-axis) vs. MetaPSICOV.

Table 6. A summary of our blind prediction results on 5 CAMEO hard targets with a new fold.

| Target | CAMEO ID | Type | Len | Meff | Method | RMSD(Å) | TMscore |
|---|---|---|---|---|---|---|---|
| 2nc8A | 2016-09-10_00000002_1 | β | 182 | 250 | Our server | 6.5 | 0.61 |
| | | | | | Best of the others | 12.18 | 0.47 |
| 5dcjA | 2016-09-17_00000018_1 | α+β | 125 | 180 | Our server | 7.9 | 0.52 |
| | | | | | Best of the others | 10.0 | 0.53 |
| 5djeB | 2016-09-24_00000052_1 | α | 140 | 330 | Our server | 5.81 | 0.65 |
| | | | | | Best of the others | 14.98 | 0.34 |
| 5f5pH | 2016-10-15_00000047_1 | α | 217 | 65 | Our server | 4.21 | 0.71 |

| | | | | | Best of the others | >40.0 | 0.48 |
| --- | --- | --- | --- | --- | --- | --- | --- |
| 5flgB | 2016-11-12_00000046_1 | α/β | 260 | 113 | Our server | 7.12 | 0.61 |
| | | | | | Best of the others | 16.9 | 0.25 |

Among these 41 hard targets, there are five multi-domain proteins: 5idoA, 5hmqF, 5b86B, 5b2gG and 5cylH. Table 7 shows that the average contact prediction accuracy of our method on these 5 multi-domain proteins is much better than the others. For multi-domain proteins, we use a superposition-independent score lDDT instead of TMscore to measure the quality of a 3D model. As shown in Table 8, the 3D models built by our server from predicted contacts have much better lDDT score than those built from CCMpred and MetaPSICOV.

**Table 7.** The average contact prediction accuracy of our method and the others on 5 multi-domain proteins among the 41 CAMEO hard targets.

| Method | Short | | | | Medium | | | | Long | | | |
| --- | --- | --- | --- | --- | --- | --- | --- | --- | --- | --- | --- | --- |
| | L/10 | L/5 | L/2 | L | L/10 | L/5 | L/2 | L | L/10 | L/5 | L/2 | L |
| EVfold | 0.17 | 0.13 | 0.09 | 0.07 | 0.18 | 0.12 | 0.08 | 0.06 | 0.54 | 0.40 | 0.26 | 0.18 |
| PSICOV | 0.27 | 0.18 | 0.10 | 0.07 | 0.26 | 0.17 | 0.11 | 0.07 | 0.62 | 0.49 | 0.31 | 0.20 |
| plmDCA | 0.29 | 0.23 | 0.11 | 0.07 | 0.32 | 0.22 | 0.11 | 0.08 | 0.66 | 0.51 | 0.34 | 0.22 |
| Gremlin | 0.30 | 0.24 | 0.12 | 0.08 | 0.32 | 0.22 | 0.12 | 0.07 | 0.67 | 0.52 | 0.36 | 0.23 |
| CCMpred | 0.30 | 0.23 | 0.12 | 0.08 | 0.32 | 0.22 | 0.12 | 0.08 | 0.66 | 0.51 | 0.35 | 0.23 |
| MetaPSICOV | 0.52 | 0.37 | 0.21 | 0.14 | 0.32 | 0.26 | 0.16 | 0.11 | 0.72 | 0.58 | 0.41 | 0.26 |
| Our method | 0.74 | 0.58 | 0.33 | 0.19 | 0.68 | 0.55 | 0.33 | 0.20 | 0.96 | 0.91 | 0.76 | 0.57 |

**Table 8.** The lDDT score of the 3D models built for the 5 multi-domain proteins using predicted contacts.

| Targets | Length | CCMpred | MetaPSICOV | Our |
| --- | --- | --- | --- | --- |
| 5idoA | 512 | 23.67 | 24.24 | 36.83 |
| 5hmqF | 637 | 24.84 | 25.91 | 33.16 |
| 5b86B | 600 | 29.88 | 32.85 | 42.58 |
| 5b2gG | 364 | 28.52 | 30.47 | 47.91 |
| 5cylH | 370 | 22.21 | 23.37 | 30.62 |

### Study of CAMEO target 2nc8A (CAMEO ID: 2016-09-10_00000002_1, PDB ID:2nc8)

On September 10, 2016, CAMEO released two hard test targets for structure prediction. Our contact server successfully folded the hardest one (PDB ID: 2nc8), a mainly β protein of 182 residues. Table 9 shows that our server produced a much better contact prediction than CCMpred and MetaPSICOV. CCMpred has very low accuracy since HHblits detected only ~250 non-redundant sequence homologs for this protein, i.e., its Meff=250. Fig. 6 shows the predicted contact maps and their overlap with the native. MetaPSICOV fails to predict many long-range contacts while CCMpred introduces too many false positives.

**Table 9.** The long- and medium-range contact prediction accuracy of our method, MetaPSICOV and CCMpred on the CAMEO target 2nc8A.

|  | Long-range accuracy | | | | Medium-range accuracy | | | |
| --- | --- | --- | --- | --- | --- | --- | --- | --- |
|  | L | L/2 | L/5 | L/10 | L | L/2 | L/5 | L/10 |
| Our method | 0.764 | 0.923 | 0.972 | 1.0 | 0.450 | 0.769 | 0.972 | 1.0 |
| MetaPSICOV | 0.258 | 0.374 | 0.556 | 0.667 | 0.390 | 0.626 | 0.806 | 0.944 |
| CCMpred | 0.165 | 0.231 | 0.389 | 0.333 | 0.148 | 0.187 | 0.167 | 0.222 |

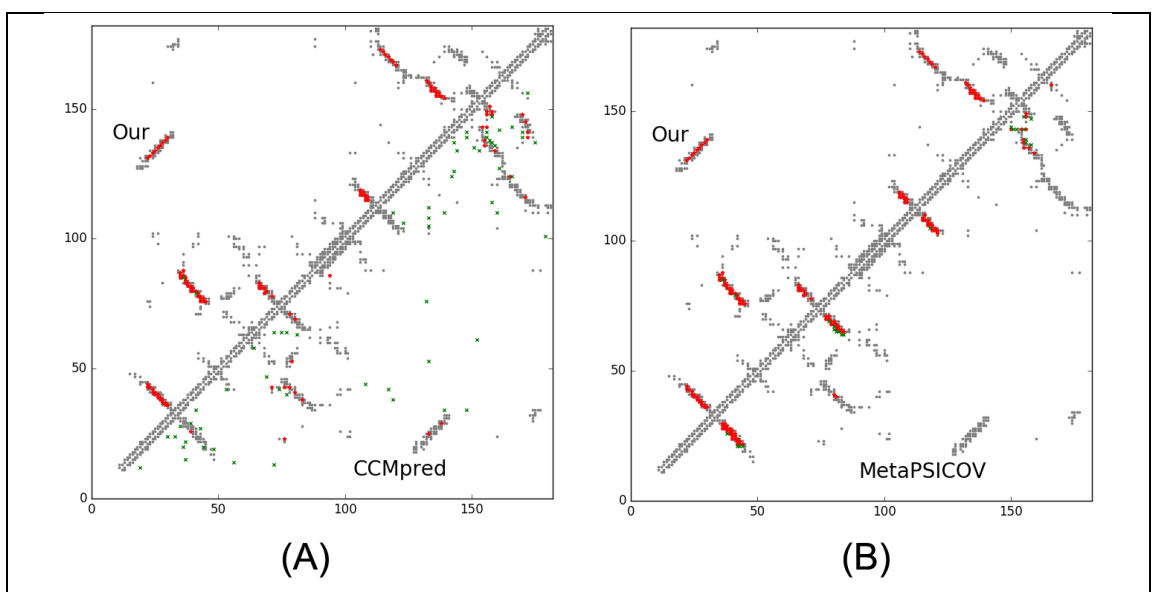

**Figure 6.** Overlap between top L/2 predicted contacts (in red or green) and the native (in grey). Red (green) dots indicate correct (incorrect) prediction. The left picture shows the comparison between our prediction (in upper-left triangle) and CCMpred (in lower-right triangle) and the right picture shows the comparison between our prediction (in upper-left triangle) and MetaPSICOV (in lower-right triangle).

The 3D model submitted by our contact server has TMscore 0.570 (As of September 16, 2016, our server submits only one 3D model for each test protein) and the best of our top 5 models has TMscore 0.612 and RMSD 6.5Å. Fig. 7 shows that the beta strands of our predicted model (red) matches well with the native (blue). To examine the superimposition of our model with its native structure from various angles, please see http://raptorx.uchicago.edu/DeepAlign/75097011/. By contrast, the best of top 5 models built by CNS from CCMpred- and MetaPSICOV-predicted contacts have TMscore 0.206 and 0.307, respectively, and RMSD 15.8Å and 14.2Å, respectively. The best TMscore obtained by the other CAMEO-participating servers is only 0.47 (Fig. 8). Three top-notch servers HHpred, RaptorX and Robetta only submitted models with TMscore≤0.30. According to Xu and Zhang (39), a 3D model with TMscore<0.5 is unlikely to have a correct fold while a model with TMscore≥0.6 surely has a correct fold. That is, our contact server predicted a correct fold for this test protein while the others failed to.

This test protein represents almost a novel fold. Our in-house structural homolog search tool DeepSearch(40) cannot identify structurally very similar proteins in PDB70 (created right before September 10, 2016) for this test protein. PDB70 is a set of representative structures in PDB, in which any two share less than 70% sequence identity. DeepSearch returned two weakly similar proteins 4kx7A and 4g2aA, which have TMscore 0.521 and 0.535 with the native structure of the test protein, respectively, and TMscore 0.465 and 0.466 with our best model, respectively. This is consistent with the fact that none of the template-based servers in CAMEO submitted a model with TMscore>0.5. We cannot find structurally similar proteins in PDB70 for our best model either; the best TMscore between PDB70 and our best model is only 0.480. That is, the models predicted by our method are not simply copied from the solved structures in PDB, and our method can indeed fold a relatively large β protein with a novel fold.

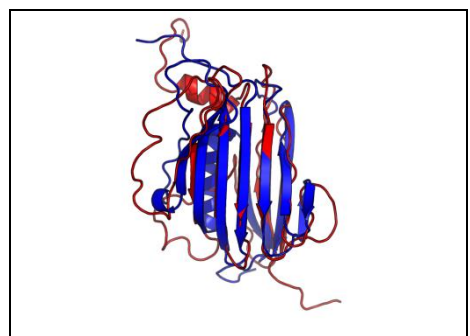

**Figure 7.** Superimposition between our predicted model (red) and its native structure (blue) for the CAMEO test protein (PDB ID 2nc8 and chain A).

| Server Name | Predictions | Resp. time (hh:mm:ss) | From | To | Cov. (%) | lDDT | lDDT Cα | Avg. lDDT-BS | Avg. lDDT-BS details | QScore | QScore details | CAD-Score | GDT_HA | RMSD | GDC | Model Conf. | MaxSub | TMScore |
|---|---|---|---|---|---|---|---|---|---|---|---|---|---|---|---|---|---|---|
| Server 60 | Model 1 | 00:51:19 | 1 | 182 | 100 | 42.76 | 50.39 | - | - | - | - | 0.48 | 26.46 | 7.69 | 36.04 | 0.50 | 0.37 | 0.57 |
| Server 56 | Model 1 | 20:53:42 | 1 | 182 | 100 | 35.81 | 43.06 | - | - | - | - | 0.43 | 19.88 | 12.18 | 27.65 | 0.80 | 0.28 | 0.47 |
| Server 58 | Model 1 | 20:54:33 | 1 | 182 | 100 | 35.81 | 43.06 | - | - | - | - | 0.43 | 19.88 | 12.18 | 27.65 | 0.80 | 0.28 | 0.47 |
| RaptorX | Model 1 | 01:17:35 | 1 | 182 | 100 | 28.73 | 32.74 | - | - | - | - | 0.41 | 12.57 | 17.55 | 16.32 | 0.65 | 0.16 | 0.30 |
| Server 57 | Model 1 | 20:54:44 | 1 | 182 | 100 | 28.64 | 33.07 | - | - | - | - | 0.39 | 12.43 | 13.54 | 18.68 | 0.73 | 0.17 | 0.36 |
| Server 45 | Model 1 | 01:51:45 | 1 | 182 | 100 | 28.45 | 32.88 | - | - | - | - | 0.43 | 19.01 | 21.83 | 22.86 | 0.65 | 0.23 | 0.36 |
| Robetta | Model 1 | 51:20:57 | 10 | 182 | 95 | 28.33 | 32.62 | - | - | - | - | 0.45 | 10.23 | 25.10 | 11.51 | 0.50 | 0.12 | 0.21 |
| HHpredB | Model 1 | 12:14:59 | 1 | 182 | 100 | 23.70 | 28.37 | - | - | - | - | 0.40 | 12.87 | 20.72 | 16.16 | 0.85 | 0.17 | 0.30 |
| Princeton_TEMPLATE | Model 1 | 01:02:52 | 1 | 182 | 100 | 23.38 | 27.09 | - | - | - | - | 0.38 | 9.94 | 23.55 | 11.52 | 0.59 | 0.12 | 0.24 |
| SPARKS-X | Model 1 | 00:12:47 | 1 | 182 | 100 | 23.08 | 26.26 | - | - | - | - | 0.37 | 7.60 | 19.12 | 8.89 | 0.52 | 0.09 | 0.20 |
| Server 55 | Model 1 | 00:28:24 | 1 | 182 | 100 | 22.38 | 25.78 | - | - | - | - | 0.39 | 7.60 | 23.65 | 7.81 | 0.67 | 0.08 | 0.20 |
| RBO Aleph | Model 1 | 65:29:29 | 1 | 182 | 100 | 21.52 | 23.78 | - | - | - | - | 0.35 | 5.99 | 20.90 | 6.86 | 0.80 | 0.07 | 0.17 |

**Figure 8.** The list of CAMEO-participating servers (only 12 of 20 are displayed) and their model scores. The rightmost column displays the TMscore of submitted models. Server60 is our contact web server.

### Study of CAMEO target 5dcjA (CAMEO ID: 2016-09-17_00000018_1, PDB ID:5dcj)

This target was released by CAMEO on September 17, 2016. It is an α+β sandwich protein of 125 residues. The four beta sheets of this protein are wrapped by one and three alpha helices at two sides. Table 10 shows that our server produced a much better contact prediction than CCMpred and MetaPSICOV. Specifically, the contact map predicted by our method has L/2 long-range accuracy 0.645 while that by CCMpred and MetaPSICOV has L/2 accuracy only 0.05 and 0.194, respectively.

CCMpred has very low accuracy since HHblits can only find ~180 non-redundant sequence homologs for this protein, i.e., its Meff=180. Fig. 9 shows the predicted contact maps and their overlap with the native. Both CCMpred and metaPSICOV failed to predict some long-range contacts.

**Table 10.** The long- and medium-range contact prediction accuracy of our method, MetaPSICOV and CCMpred on the CAMEO target 5dcjA.

|  | **Long range** | | | | **Medium range** | | | |
| --- | --- | --- | --- | --- | --- | --- | --- | --- |
|  | L | L/2 | L/5 | L/10 | L | L/2 | L/5 | L/10 |
| Our method | 0.456 | 0.645 | 0.88 | 0.833 | 0.36 | 0.645 | 0.92 | 1.0 |
| metaPSICOV | 0.144 | 0.194 | 0.32 | 0.25 | 0.344 | 0.532 | 0.8 | 1.0 |
| CCMpred | 0.05 | 0.05 | 0.08 | 0.08 | 0.1 | 0.129 | 0.12 | 0.25 |

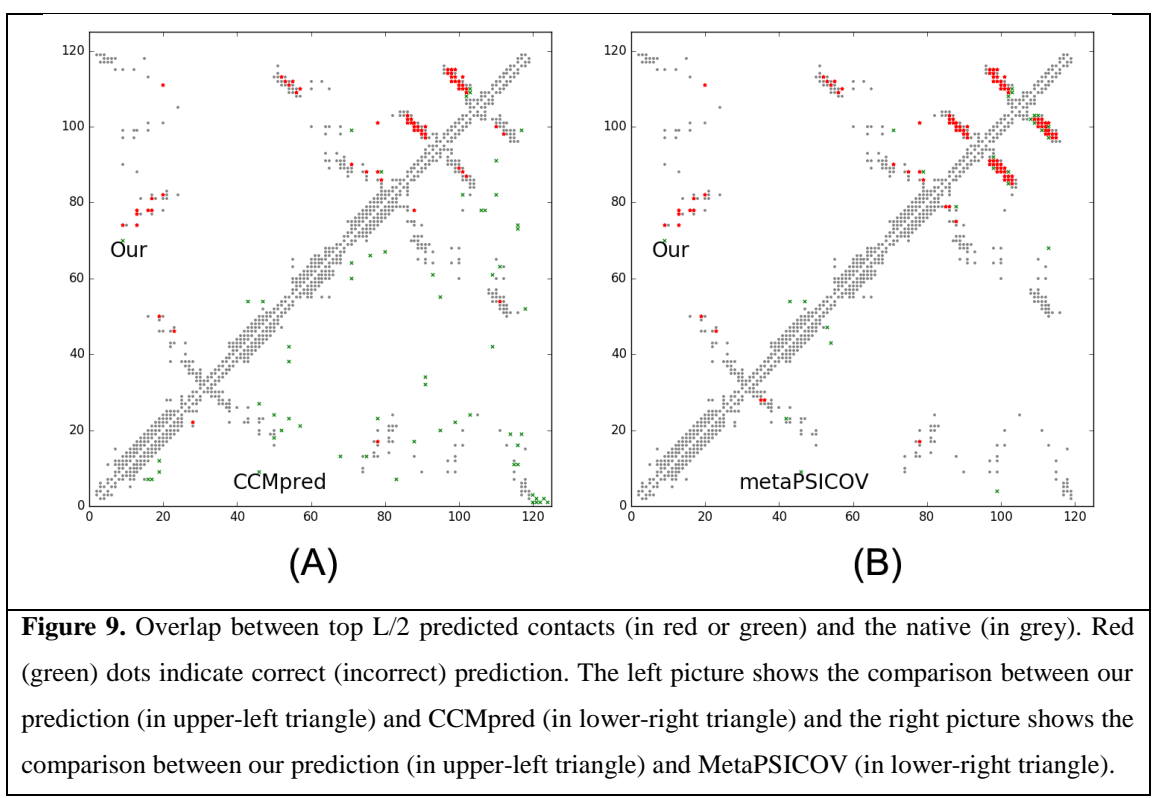

**Figure 9.** Overlap between top L/2 predicted contacts (in red or green) and the native (in grey). Red (green) dots indicate correct (incorrect) prediction. The left picture shows the comparison between our prediction (in upper-left triangle) and CCMpred (in lower-right triangle) and the right picture shows the comparison between our prediction (in upper-left triangle) and MetaPSICOV (in lower-right triangle).

The first 3D model submitted by our contact server has TMscore 0.50 and the best of our 5 models has TMscore 0.52 and RMSD 7.9Å. The best of top 5 models built by CNS from CCMpred- and MetaPSICOV-predicted contacts have TMscore 0.243 and 0.361, respectively. Fig. 10(A) shows that all the beta strands and the three surrounding alpha helices of our predicted model (in red) matches well with the native structure (blue), while the models from CCMpred (Fig.10(B)) and MetaPSICOV (Fig.10(C)) do not have a correct fold. To examine the superimposition of our model with its native structure from various angles, please see http://raptorx.uchicago.edu/DeepAlign/92913404/ .

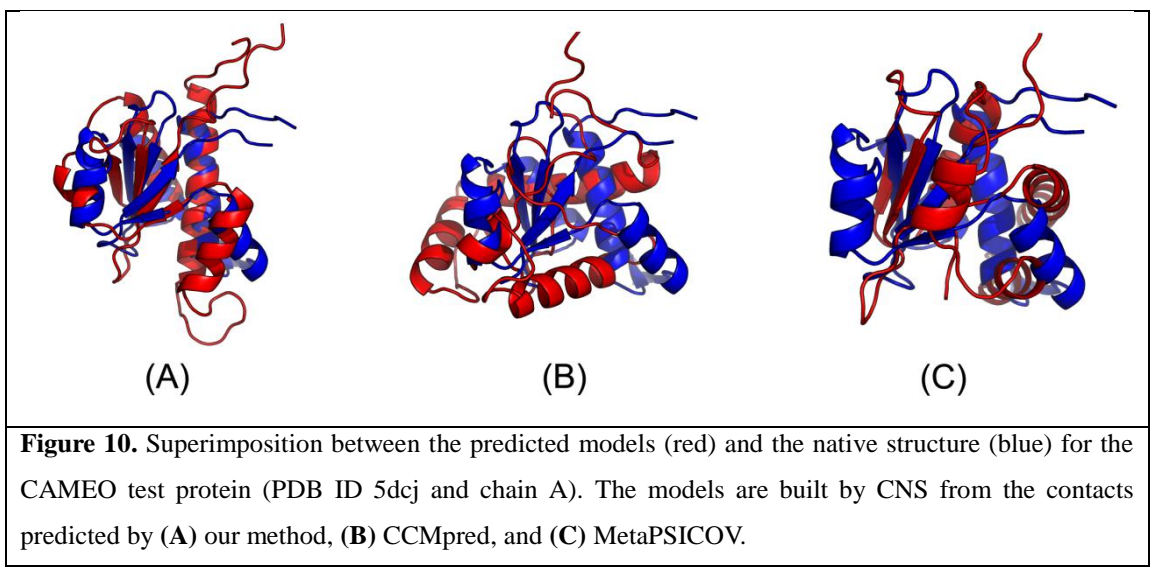

**Figure 10.** Superimposition between the predicted models (red) and the native structure (blue) for the CAMEO test protein (PDB ID 5dcj and chain A). The models are built by CNS from the contacts predicted by **(A)** our method, **(B)** CCMpred, and **(C)** MetaPSICOV.

In terms of TMscore, our models have comparable quality to Robetta, but better than the other servers (Fig. 11). In terms of lDDT-Cα score, our models are better than all the others. In particular, our method produced better models than the popular homology modeling server HHpredB and our own template-based modeling server RaptorX, which submitted models with TMscore≤0.45.

This test protein represents a novel fold. Searching through PDB70 created right before September 17, 2016 by our in-house structural homolog search tool DeepSearch cannot identify structurally similar proteins for this test protein. The most structurally similar proteins are 3lr5A and 5ereA, which have TMscore 0.431 and 0.45 with the test protein, respectively. This is consistent with the fact that none of the template-based servers in CAMEO can predict a good model for this test protein. By contrast, our contact-assisted model has TMscore 0.52, which is higher than all the template-based models.

| Server Name | Predictions | Resp. time (hh:mm:ss) | From | To | Cov. (%) | lDDT | lDDT Cα | Avg. lDDT-BS | Avg. lDDT-BS details | QScore | QScore details | CAD-Score | GDT_HA | RMSD | GDC | Model Conf. | MaxSub | TMScore |
|---|---|---|---|---|---|---|---|---|---|---|---|---|---|---|---|---|---|---|
| Server 60 | Model 1 | 11:38:06 | 1 | 125 | 100 | 47.88 | 57.13 | 43.45 | CPS1:0.43(1.00) | - | - | 0.51 | 27.97 | 8.93 | 32.76 | 0.50 | 0.35 | 0.50 |
| Robetta | Model 1 | 11:58:59 | 1 | 125 | 100 | 48.12 | 54.58 | 49.79 | CPS1:0.50(1.00) | - | - | 0.53 | 29.66 | 10.39 | 36.80 | 0.90 | 0.41 | 0.50 |
| Server 56 | Model 1 | 21:07:50 | 1 | 125 | 100 | 46.12 | 53.12 | 39.74 | CPS1:0.40(1.00) | - | - | 0.51 | 28.39 | 10.06 | 34.81 | 0.96 | 0.38 | 0.48 |
| Server 58 | Model 1 | 21:06:20 | 1 | 125 | 100 | 46.12 | 53.12 | 39.74 | CPS1:0.40(1.00) | - | - | 0.51 | 28.39 | 10.06 | 34.81 | 0.96 | 0.38 | 0.48 |
| RaptorX | Model 1 | 10:28:22 | 1 | 125 | 100 | 45.12 | 50.42 | 38.20 | CPS1:0.38(1.00) | - | - | 0.51 | 26.91 | 10.10 | 32.71 | 0.65 | 0.32 | 0.45 |
| Princeton_TEMPLATE | Model 1 | 04:55:57 | 1 | 125 | 100 | 44.32 | 50.33 | 37.68 | CPS1:0.38(1.00) | - | - | 0.47 | 23.73 | 10.69 | 31.45 | 0.50 | 0.33 | 0.45 |
| Server 45 | Model 1 | 10:53:53 | 1 | 125 | 100 | 44.39 | 49.91 | 35.88 | CPS1:0.36(1.00) | - | - | 0.51 | 26.70 | 11.97 | 33.12 | 0.64 | 0.34 | 0.46 |
| SPARKS-X | Model 1 | 00:46:54 | 1 | 125 | 100 | 42.67 | 49.20 | 36.24 | CPS1:0.36(1.00) | - | - | 0.49 | 25.64 | 11.71 | 32.24 | 0.54 | 0.33 | 0.45 |
| HHpredB | Model 1 | 80:54:59 | 1 | 125 | 100 | 42.56 | 48.88 | 37.32 | CPS1:0.37(1.00) | - | - | 0.49 | 26.27 | 11.62 | 32.21 | 0.89 | 0.33 | 0.45 |
| Server 55 | Model 1 | 00:08:10 | 1 | 125 | 100 | 42.14 | 48.44 | 36.60 | CPS1:0.37(1.00) | - | - | 0.50 | 26.27 | 10.16 | 31.85 | 0.88 | 0.33 | 0.45 |
| Server 54 | Model 1 | 00:00:57 | 3 | 121 | 95 | 42.29 | 48.43 | 37.01 | CPS1:0.37(1.00) | - | - | 0.50 | 26.48 | 10.13 | 31.75 | 0.89 | 0.33 | 0.45 |
| SWISS-MODEL | Model 1 | 00:01:06 | 3 | 121 | 95 | 41.93 | 48.31 | 34.95 | CPS1:0.35(1.00) | - | - | 0.49 | 27.33 | 10.15 | 31.78 | 0.90 | 0.33 | 0.45 |
| Server 48 | Model 1 | 00:02:00 | 3 | 121 | 95 | 41.93 | 48.31 | 34.90 | CPS1:0.35(1.00) | - | - | 0.49 | 27.33 | 10.15 | 31.78 | 0.90 | 0.33 | 0.45 |
| IntFOLD3-TS | Model 1 | 22:08:20 | 1 | 125 | 100 | 42.67 | 48.00 | 38.51 | CPS1:0.39(1.00) | - | - | 0.47 | 25.42 | 11.87 | 30.96 | 0.74 | 0.31 | 0.44 |

**Figure 11.** The list of CAMEO-participating servers (only 14 of 20 are displayed) and their model scores, sorted by lDDT-Cα. The rightmost column displays the TMscore of submitted models. Server60 is our contact web server.

## Study of CAMEO target 5djeB (CAMEO ID: 2016-09-24_00000052_1, PDB ID: 5dje)

This target was released on September 24, 2016. It is an alpha protein of 140 residues with a novel fold. Table 11 shows that our server produced a much better contact prediction than CCMpred and MetaPSICOV. Specifically, the contact map predicted by our method has L/5 and L/10 long-range accuracy 50.0% and 71.4%, respectively, while that by CCMpred and MetaPSICOV has L/5 and L/10 accuracy less than 30%. CCMpred has low accuracy since HHblits can only find ~330 non-redundant sequence homologs for this protein, i.e., its Meff=330. Fig. 12 shows the predicted contact maps and their overlap with the native. Both CCMpred and metaPSICOV failed to predict some long-range contacts.

**Table 11.** The long- and medium-range contact prediction accuracy of our method, MetaPSICOV and CCMpred on the CAMEO target 5djeB.

|  | Long range accuracy | | | | Medium range accuracy | | | |
|---|---|---|---|---|---|---|---|---|
|  | L | L/2 | L/5 | L/10 | L | L/2 | L/5 | L/10 |
| Our method | 0.300 | 0.357 | 0.500 | 0.714 | 0.186 | 0.229 | 0.357 | 0.357 |
| metaPSICOV | 0.193 | 0.200 | 0.286 | 0.286 | 0.100 | 0.143 | 0.214 | 0.286 |
| CCMpred | 0.079 | 0.114 | 0.107 | 0.214 | 0.036 | 0.029 | 0.071 | 0.143 |

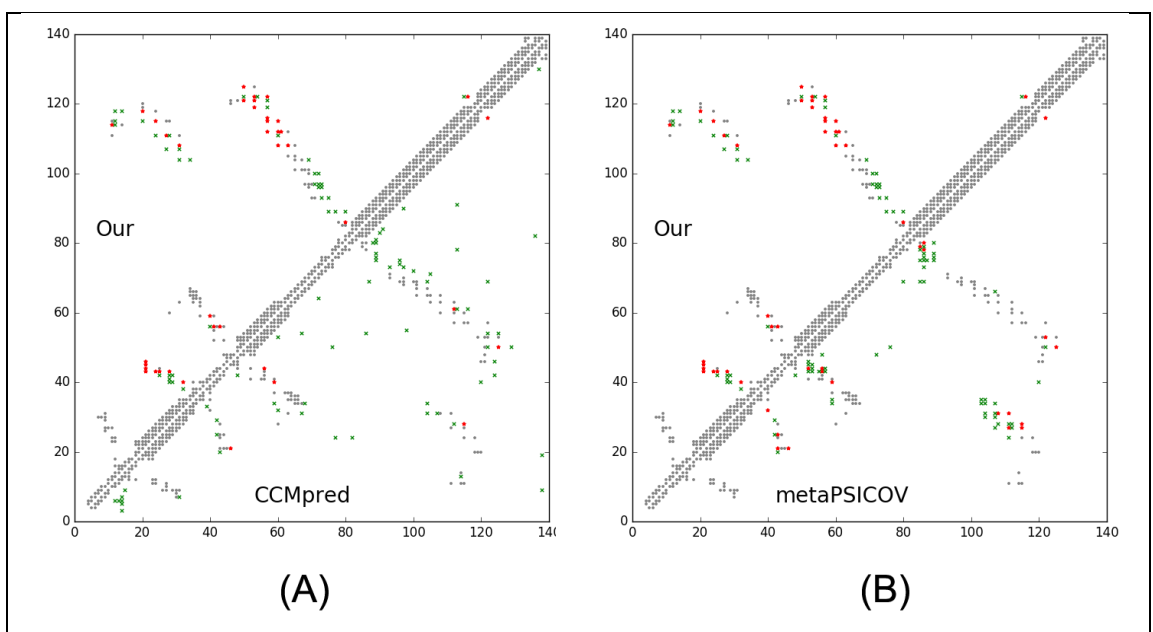

**Figure 12.** Overlap between top L/2 predicted contacts (in red and green) and the native (in grey). Red (green) dots indicate correct (incorrect) prediction. The left picture shows the comparison between our prediction (in upper-left triangle) and CCMpred (in lower-right triangle) and the right picture shows the comparison between our prediction (in upper-left triangle) and MetaPSICOV (in lower-right triangle).

The first 3D model submitted by our contact server has TMscore 0.65, while the best of our 5 models has TMscore 0.65 and RMSD 5.6Å. By contrast, the best of top 5 models built by CNS from CCMpred- and MetaPSICOV-predicted contacts have TMscore 0.404 and 0.427, respectively. Fig. 13(A) shows that all the four alpha helices of our predicted model (in red) matches well with the native structure (blue), while the models from CCMpred (Fig. 13(B)) and MetaPSICOV (Fig. 13(C)) fail to predict the 3rd long helix correctly. To examine the superimposition of our model with its native structure from various angles, please see http://raptorx.uchicago.edu/DeepAlign/26652330/. Further, all other CAMEO registered servers, including the top-notch servers such as HHpred, RaptorX, SPARKS-X, and RBO Aleph (template-based and ab initio folding) only submitted models with TMscore≤0.35, i.e., failed to predict a correct fold (Fig. 14).

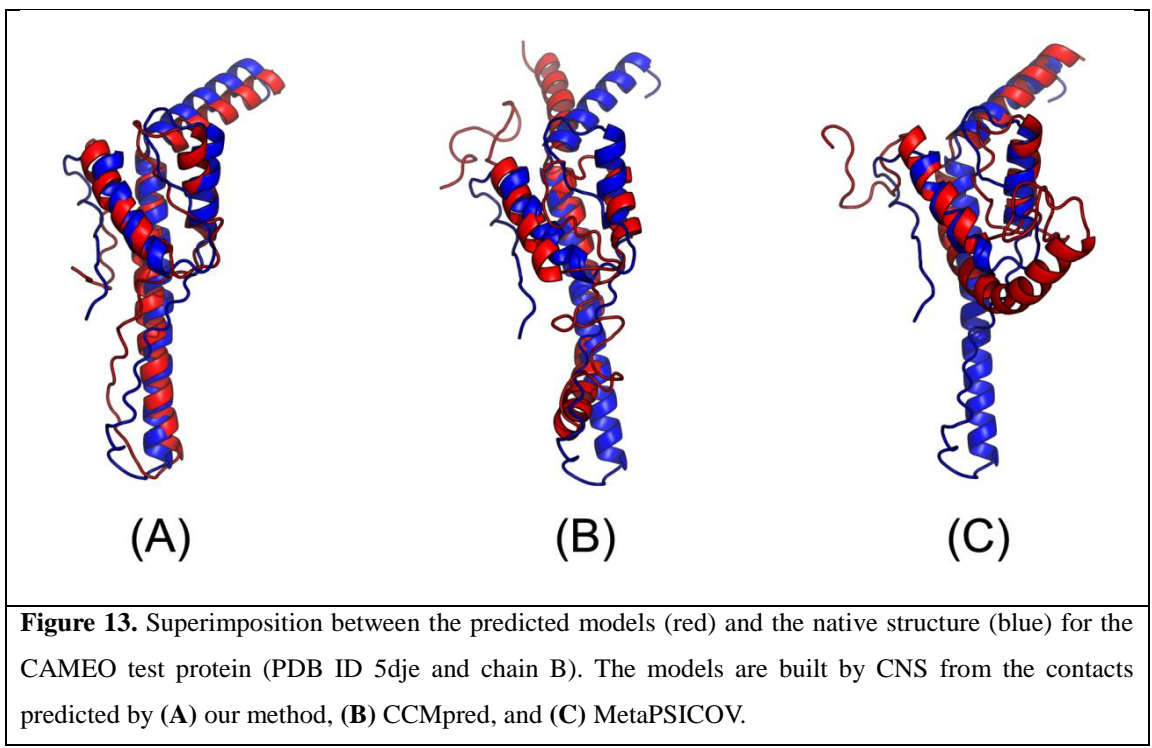

**Figure 13.** Superimposition between the predicted models (red) and the native structure (blue) for the CAMEO test protein (PDB ID 5dje and chain B). The models are built by CNS from the contacts predicted by **(A)** our method, **(B)** CCMpred, and **(C)** MetaPSICOV.

This test protein represents a novel fold. Searching through PDB70 created right before September 24, 2016 by our in-house structural homolog search tool DeepSearch cannot identify structurally similar proteins for this test protein. The most structurally similar proteins are 1u7lA and 4x5uA, which have TMscore 0.439 and 0.442 with the test protein, respectively. This is consistent with the fact that none of the template-based CAMEO-participating servers predicted a good model for this test protein. By contrast, our contact-assisted model has TMscore 0.65, much better than all the template-based models.

| Server Name | Predictions | Resp. time (hh:mm:ss) | From | To | Cov. (%) | lDDT | lDDT Cα | Avg. lDDT-BS | Avg. lDDT-BS details | QScore | QScore details | CAD-Score | GDT_HA | RMSD | GDC | Model Conf. | MaxSub | TMScore |
|---|---|---|---|---|---|---|---|---|---|---|---|---|---|---|---|---|---|---|
| Server 60 | Model 1 | 19:20:14 | 1 | 140 | 100 | 54.43 | 68.22 | - | | - | | 0.60 | 36.03 | 5.81 | 45.19 | 0.50 | 0.51 | 0.65 |
| RaptorX | Model 1 | 14:50:24 | 1 | 140 | 100 | 33.59 | 42.23 | - | | - | | 0.53 | 18.93 | 18.88 | 19.98 | 0.71 | 0.22 | 0.34 |
| RBO Aleph | Model 1 | 53:34:20 | 1 | 140 | 100 | 40.85 | 50.35 | - | | - | | 0.56 | 18.57 | 14.98 | 18.37 | 0.50 | 0.19 | 0.31 |
| Server 45 | Model 1 | 36:00:52 | 1 | 140 | 100 | 35.96 | 44.10 | - | | - | | 0.53 | 19.12 | 20.97 | 20.16 | 0.67 | 0.21 | 0.31 |
| SPARKS-X | Model 1 | 01:34:53 | 1 | 140 | 100 | 34.24 | 43.59 | - | | - | | 0.51 | 16.54 | 24.40 | 15.18 | 0.53 | 0.19 | 0.27 |
| IntFOLD3-TS | Model 1 | 23:58:21 | 1 | 140 | 100 | 33.49 | 41.34 | - | | - | | 0.49 | 16.91 | 25.85 | 14.94 | 0.75 | 0.18 | 0.26 |
| Princeton_TEMPLATE | Model 1 | 05:43:56 | 1 | 140 | 100 | 35.77 | 44.78 | - | | - | | 0.53 | 20.22 | 23.15 | 15.79 | 0.50 | 0.21 | 0.26 |
| Server 56 | Model 1 | 22:29:49 | 1 | 140 | 100 | 34.96 | 43.55 | - | | - | | 0.48 | 15.44 | 23.61 | 14.73 | 0.96 | 0.17 | 0.26 |
| Server 58 | Model 1 | 22:29:22 | 1 | 140 | 100 | 34.96 | 43.55 | - | | - | | 0.48 | 15.44 | 23.61 | 14.73 | 0.96 | 0.17 | 0.26 |
| Server 57 | Model 1 | 22:29:51 | 1 | 140 | 100 | 34.09 | 42.06 | - | | - | | 0.50 | 16.54 | 26.14 | 14.42 | 0.82 | 0.18 | 0.25 |
| Server 7 | Model 1 | 18:14:55 | 81 | 140 | 42 | 15.51 | 20.19 | - | | - | | 0.25 | 16.18 | 14.71 | 15.77 | 0.58 | 0.19 | 0.22 |
| HHpredB | Model 1 | 00:08:59 | 1 | 140 | 100 | 32.19 | 41.28 | - | | - | | 0.53 | 14.71 | 36.79 | 12.95 | 0.61 | 0.15 | 0.21 |
| Server 55 | Model 1 | 00:35:17 | 1 | 140 | 100 | 30.19 | 37.98 | - | | - | | 0.53 | 16.36 | 35.62 | 14.87 | 0.66 | 0.17 | 0.21 |
| SWISS-MODEL | Model 1 | 00:00:14 | 86 | 127 | 30 | 9.27 | 12.47 | - | | - | | 0.17 | 16.18 | 11.45 | 14.77 | 0.92 | 0.17 | 0.19 |
| Server 46 | Model 1 | 00:03:45 | 86 | 127 | 30 | 9.27 | 12.47 | - | | - | | 0.17 | 16.18 | 11.45 | 14.77 | 0.92 | 0.17 | 0.19 |

**Figure 14.** The list of CAMEO-participating servers (only 15 of 20 are displayed) and their model scores. The rightmost column displays the TMscore of submitted models. Server60 is our contact web server.

**Study of CAMEO target 5f5pH (CAMEO ID: 2016-10-15_00000047_1, PDB ID: 5f5p)**

On October 15, 2016, our contact web server successfully folded a very hard and also interesting CAMEO target (PDB ID: 5f5pH, CAMEO ID: 2016-10-15_00000047_1). This target is an alpha protein of 217 residues with four helices. Table 12 shows that our server produced a much better long-range contact prediction than CCMpred and MetaPSICOV. Specifically, our contact prediction has L/5 and L/10 long-range accuracy 76.7% and 95.2%, respectively, while MetaPSICOV has L/5 and L/10 accuracy less than 40%. CCMpred has very low accuracy since this target has only ~65 non-redundant sequence homologs, i.e., its Meff=65. The three methods have low L/k (k=1, 2) medium-range accuracy because there are fewer than L/k native medium-range contacts while we use L/k as the denominator in calculating accuracy. As shown in Fig. 15, CCMpred predicts too many false positives while MetaPSICOV predicts very few correct long-range contacts.

**Table 12.** The long- and medium-range contact prediction accuracy of our method, MetaPSICOV and CCMpred on the CAMEO target 5f5pH.

|  | Long-range accuracy | | | | Medium-range accuracy | | | |
|---|---|---|---|---|---|---|---|---|
|  | L | L/2 | L/5 | L/10 | L | L/2 | L/5 | L/10 |
| Our server | 0.382 | 0.602 | 0.767 | 0.952 | 0.041 | 0.083 | 0.209 | 0.381 |
| metaPSICOV | 0.161 | 0.250 | 0.326 | 0.476 | 0.041 | 0.083 | 0.163 | 0.190 |
| CCMpred | 0.032 | 0.037 | 0.047 | 0.048 | 0.009 | 0.019 | 0.023 | 0.032 |

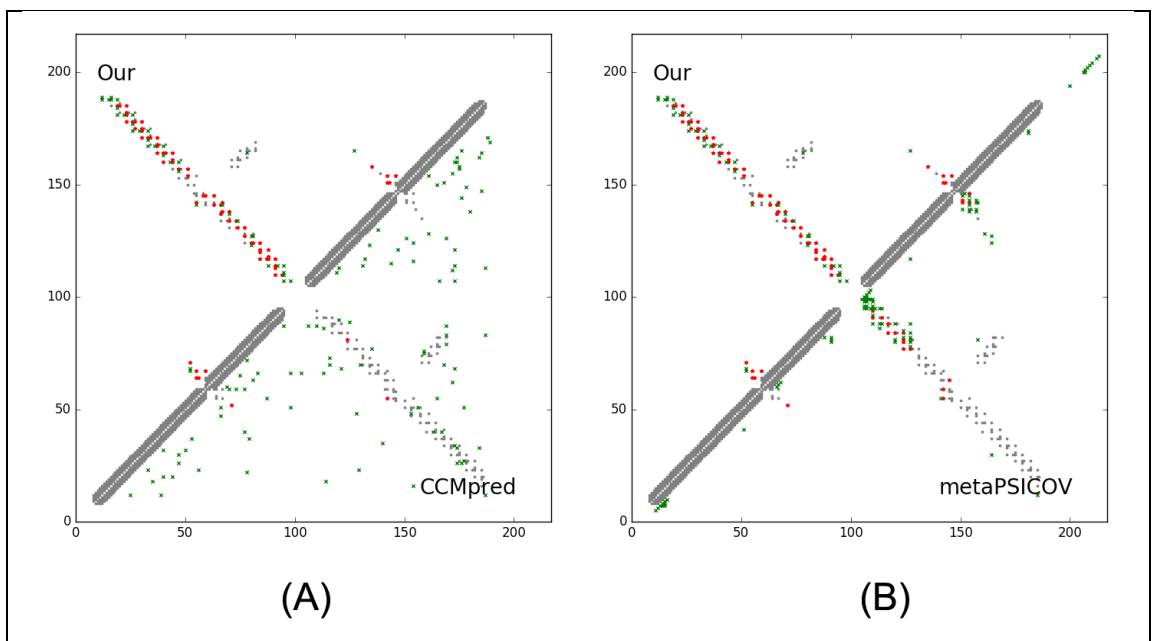

**Figure 15.** Overlap between top L/2 predicted contacts (in red and green) and the native (in grey). Red (green) dots indicate correct (incorrect) prediction. The left picture shows the comparison between our prediction (in upper-left triangle) and CCMpred (in lower-right triangle) and the right picture shows the comparison between our prediction (in upper-left triangle) and MetaPSICOV (in lower-right triangle).

Our submitted 3D model has TMscore 0.71 and RMSD 4.21Å. By contrast, the best of top 5 models built by CNS from CCMpred- and MetaPSICOV-predicted contacts have TMscore 0.280 and 0.472, respectively. Fig. 16(A) shows that our predicted model (in red) match well with the native structure (blue), while the model from CCMpred (Fig. 16(B)) is completely wrong and the model from MetaPSICOV (Fig. 16(C)) fails to place the $1^{st}$ and $4^{th}$ helices correctly. Please see http://raptorx.uchicago.edu/DeepAlign/14544627/ for the animated superimposition of our model with its native structure. As shown in the ranking list (Fig. 17), all the other CAMEO-participating servers, including Robetta, HHpred, RaptorX, SPARKS-X, and RBO Aleph (template-based and ab initio folding) only submitted models with TMscore≤0.48 and RMSD>43.82Å. Our contact server is the only one that predicted a correct fold for this target.

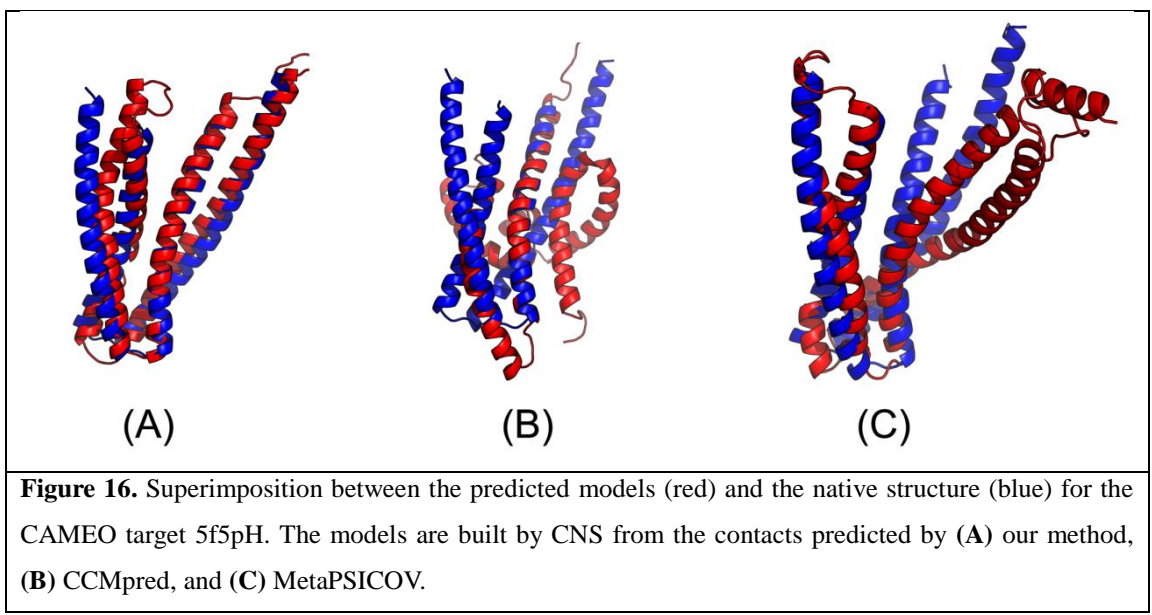

**Figure 16.** Superimposition between the predicted models (red) and the native structure (blue) for the CAMEO target 5f5pH. The models are built by CNS from the contacts predicted by **(A)** our method, **(B)** CCMpred, and **(C)** MetaPSICOV.

| Server Name | Predictions | Resp. time (hh:mm:ss) | From | To | Cov. (%) | lDDT | lDDT Cα | Avg. lDDT-BS | Avg. lDDT-BS details | QScore | QScore details | CAD-Score | GDT_HA | RMSD | GDC | Model Conf. | MaxSub | TMScore |
|---|---|---|---|---|---|---|---|---|---|---|---|---|---|---|---|---|---|---|
| Server 60 | Model 1 | 08:54:54 | 1 | 217 | 100 | 62.05 | 74.62 | - | - | - | - | 0.66 | 34.80 | 4.21 | 46.75 | 0.50 | 0.51 | 0.71 |
| Server 55 | Model 1 | 04:57:59 | 1 | 217 | 100 | 46.69 | 53.25 | - | - | - | - | 0.67 | 36.40 | 49.95 | 37.59 | 0.58 | 0.43 | 0.48 |
| SWISS-MODEL | Model 1 | 00:00:55 | 10 | 189 | 82 | 46.62 | 53.11 | - | - | - | - | 0.66 | 36.55 | 50.09 | 37.53 | 0.60 | 0.43 | 0.48 |
| Server 54 | Model 1 | 04:33:36 | 10 | 189 | 82 | 46.62 | 53.11 | - | - | - | - | 0.66 | 36.55 | 50.09 | 37.53 | 0.60 | 0.43 | 0.48 |
| Server 46 | Model 1 | 03:38:25 | 10 | 189 | 82 | 46.43 | 53.18 | - | - | - | - | 0.66 | 36.99 | 50.09 | 37.55 | 0.58 | 0.44 | 0.48 |
| Server 48 | Model 1 | 00:03:54 | 10 | 189 | 82 | 46.43 | 53.18 | - | - | - | - | 0.66 | 36.99 | 50.09 | 37.55 | 0.58 | 0.44 | 0.48 |
| Server 0 | Model 1 | 00:31:28 | 10 | 189 | 82 | 45.68 | 53.07 | - | - | - | - | 0.65 | 36.70 | 50.10 | 37.30 | 0.65 | 0.43 | 0.47 |
| Phyre2 | Model 1 | 00:36:56 | 12 | 189 | 82 | 43.44 | 52.59 | - | - | - | - | 0.66 | 36.11 | 50.40 | 37.07 | 0.50 | 0.43 | 0.47 |
| Server 19 | Model 1 | 32:52:37 | 1 | 217 | 100 | 46.45 | 54.13 | - | - | - | - | 0.65 | 30.70 | 44.65 | 33.58 | 0.60 | 0.38 | 0.46 |
| RaptorX | Model 1 | 11:27:06 | 1 | 217 | 100 | 43.71 | 52.40 | - | - | - | - | 0.62 | 27.34 | 50.41 | 31.75 | 0.60 | 0.36 | 0.44 |
| Server 61 | Model 1 | 00:08:19 | 10 | 189 | 82 | 45.47 | 52.90 | - | - | - | - | 0.66 | 31.43 | 49.27 | 31.78 | 0.56 | 0.35 | 0.43 |
| Server 64 | Model 1 | 00:22:11 | 10 | 189 | 82 | 45.47 | 52.90 | - | - | - | - | 0.66 | 31.43 | 49.27 | 31.78 | 0.56 | 0.35 | 0.43 |
| Server 65 | Model 1 | 00:10:21 | 10 | 189 | 82 | 45.47 | 52.90 | - | - | - | - | 0.66 | 31.43 | 49.27 | 31.78 | 0.56 | 0.35 | 0.43 |
| Robetta | Model 1 | 22:40:39 | 1 | 217 | 100 | 45.01 | 52.84 | - | - | - | - | 0.64 | 30.41 | 43.82 | 31.63 | 0.89 | 0.35 | 0.42 |
| M4T | Model 1 | 17:10:05 | 10 | 189 | 82 | 44.45 | 52.48 | - | - | - | - | 0.64 | 24.56 | 48.17 | 27.62 | 0.52 | 0.29 | 0.39 |

**Figure 17.** The list of CAMEO-participating servers (only 15 of 20 are displayed) and their model scores. The rightmost column displays the TMscore of submitted models. Server60 is our contact web server.

To make sure our best model is not simply copied from the database of solved structures, we search our best model against PDB70 created right before October 15, 2016 using our in-house structural homolog search tool DeepSearch, which yields two weakly similar proteins 2yfaA and 4k1pA. They have TMscore 0.536 and 0.511 with our best model, respectively. This implies that our model is not simply copied from a solved structure in PDB.

We ran BLAST on this target against PDB70 and surprisingly, found one protein 3thfA with E-value 3E-16 and sequence identity 35%. In fact, 3thfA and 5f5pH are two SD2 proteins from Drosophila and Human(41), respectively. Although homologous, they adopt different conformations and oligomerizations. In particular, 3thfA is a dimer and each monomer adopts a fold consisting of three segmented anti-parallel coiled-coil(42), whereas 5f5pH is a monomer that consists of two segmented antiparallel coiled-coils(41). Superimposing the Human SD2 monomer onto the Drosophila SD2 dimer shows that the former structure was located directly in between the two structurally identical halves of the latter structure (see Fig. 18(A)). That is, if our method predicts the contacts of 5f5pH by simply copying from 3thfA, it would produce a wrong 3D model. By contrast, all the other CAMEO-participating servers produced a wrong prediction for this target by using 3thfA as the template.

Since SD2 protein may have conformational change when docking with Rock SBD protein, we check if the Drosophila SD2 monomer would change to a similar fold as the Human SD2 monomer or not. According to(41), the Human SD2 adopts a similar fold no matter whether it docks with the Rock SBD or not. According to (42), although the Drosophila SD2 dimer may have conformational change in the presence of Rock, the change only occurs in the hinge regions, but not at the adjacent identical halves. That is, even conformational change happens, the Drosophila SD2 monomer would not resemble the Human SD2 monomer (Fig. 18(B)).

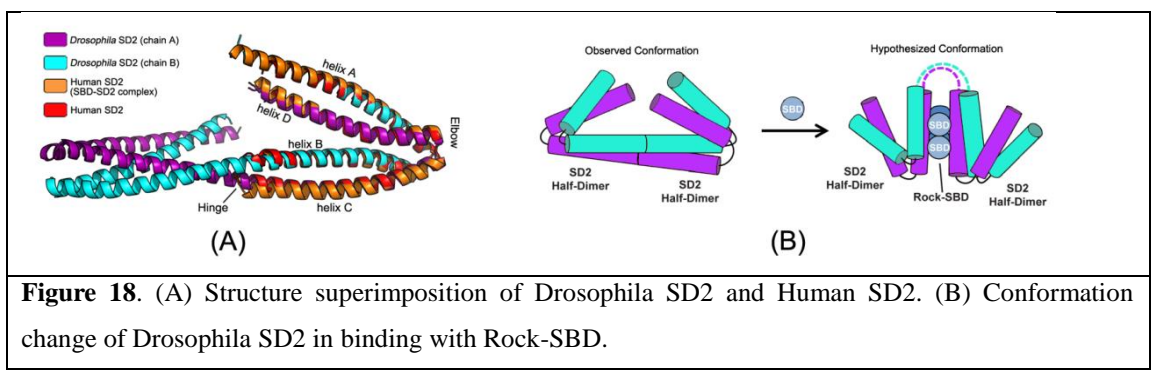

**Figure 18**. (A) Structure superimposition of Drosophila SD2 and Human SD2. (B) Conformation change of Drosophila SD2 in binding with Rock-SBD.

### Study of CAMEO target 5flgB (CAMEO ID: 2016-11-12_00000046_1, PDB ID: 5flgB)

This target was released by CAMEO on November 12, 2016 and not included in the abovementioned 41 CAMEO hard targets. This target is a unique α/β protein with 260 residues. Table 13 shows that our server produced a much better (long-range) contact prediction than CCMpred and MetaPSICOV. In particular, our predicted contact map has L, L/2, L/5 and L/10 long-range accuracy 71.1%, 86.1%, 96.1% and 100.0%, respectively, while CCMpred- and MetaPSICOV-predicted contacts have long-range accuracy less than 35% since there are only ~113 effective sequence homologs for this protein, i.e., its Meff=113. Fig. 19 shows that both CCMpred and MetaPSICOV generated many false positive contact predictions and failed to predict long-range contacts.

**Table 13.** The long- and medium-range contact prediction accuracy of our method, MetaPSICOV and CCMpred on the CAMEO target 5flgB.

|            | **Long-range accuracy** | | | | **Medium-range accuracy** | | | |
|------------|-------|-------|-------|-------|-------|-------|-------|-------|
|            | L     | L/2   | L/5   | L/10  | L     | L/2   | L/5   | L/10  |
| Our server | 0.711 | 0.861 | 0.961 | 1.00  | 0.331 | 0.500 | 0.750 | 0.808 |
| MetaPSICOV | 0.208 | 0.262 | 0.269 | 0.288 | 0.242 | 0.285 | 0.442 | 0.615 |
| CCMpred | 0.165 | 0.184 | 0.308 | 0.346 | 0.150 | 0.215 | 0.346 | 0.385 |

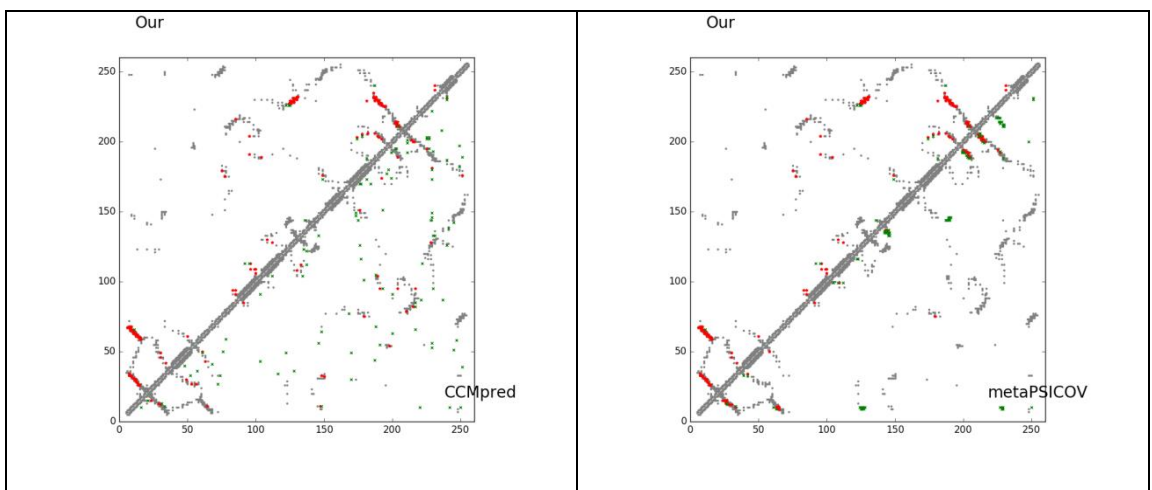

**Figure 19.** Overlap between predicted contacts (in red and green) and the native (in grey). Red (green) dots indicate correct (incorrect) prediction. Top L/2 predicted contacts by each method are shown. The left picture shows the comparison between our prediction (in upper-left triangle) and CCMpred (in lower-right triangle) and the right picture shows the comparison between our prediction (in upper-left triangle) and MetaPSICOV (in lower-right triangle).

The 3D model submitted by our contact server has TMscore 0.61 and RMSD 7.12Å. The best of top 5 models built by CNS from CCMpred- and MetaPSICOV-predicted contacts have TMscore 0.240 and 0.267, respectively. Fig. 20 shows that our method correctly modeled the overall fold, while CCMpred and MetaPSICOV failed. To examine the superimposition of our model with its native structure from various angles, please see http://raptorx.uchicago.edu/DeepAlign/12043612/. Furthermore, all the other CAMEO-participating servers, including the top-notch servers Robetta, HHpred, RaptorX, SPARKS-X, and RBO Aleph (template-based and ab initio folding), only submitted models with TMscore≤0.25 and RMSD>16.90Å (Fig. 21). A 3D model with TMscore less than 0.25 does not have the correct fold while a model with TMscore≥0.6 very likely has a correct fold. That is, our contact server predicted a correct fold for this target while the others failed to.

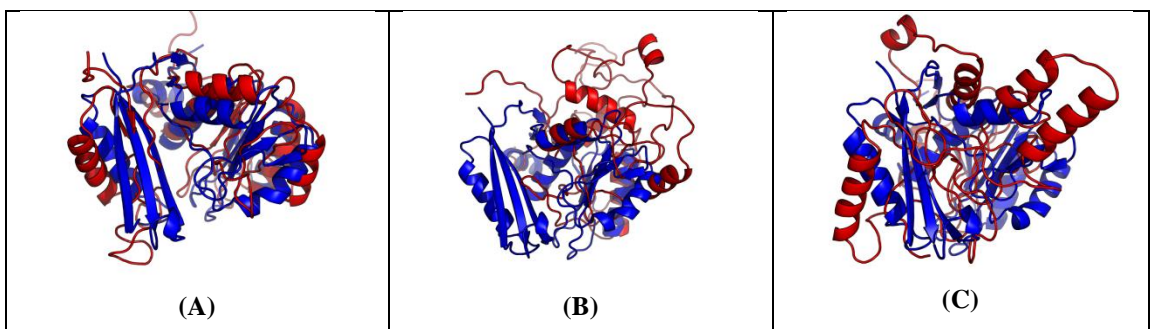

**(A)** **(B)** **(C)**

**Figure 20.** Superimposition between the predicted models (red) and the native structure (blue) for the CAMEO test protein 5flgB. The models are built by CNS from the contacts predicted by **(A)** our method, **(B)** CCMpred, and **(C)** MetaPSICOV.

This test protein has a novel fold. Searching through PDB70 created right before November 12, 2016 by our in-house structural homolog search tool DeepSearch cannot identify any similar structures. The most structurally similar proteins returned by DeepSearch are 2fb5A and 5dwmA, which have TMscore 0.367 and 0.355 with the native structure of this target, respectively. This is consistent with the fact that all the other CAMEO-participating servers failed to predict a correct fold for this target.

| Server Name | Predictions | Resp. time (hh:mm:ss) | From | To | Cov. (%) | IDDT | IDDT Cα | Avg. IDDT-BS | Avg. IDDT-BS details | QScore | QScore details | CAD-Score | GDT_HA | RMSD | GDC | Model Conf. | MaxSub | TMscore |
|---|---|---|---|---|---|---|---|---|---|---|---|---|---|---|---|---|---|---|
| Server 60 | Model 1 | 23:01:01 | 1 | 260 | 100 | 40.26 | 48.52 | 49.57 | PML5:0.39(1.00) PML1:0.40(1.00) MG4:0.78(1.00) ANP2:0.33(1.00) ANP6:0.29(1.00) MG8:0.77(1.00) | - | - | 0.46 | 21.53 | 7.12 | 32.36 | 0.50 | 0.30 | 0.61 |
| Server 56 | Model 1 | 25:20:28 | 1 | 260 | 100 | 23.25 | 28.04 | 42.32 | PML5:0.27(1.00) PML1:0.27(1.00) MG4:0.77(1.00) ANP2:0.24(1.00) ANP6:0.23(1.00) MG8:0.75(1.00) | - | - | 0.37 | 7.14 | 22.73 | 8.39 | 0.75 | 0.07 | 0.24 |
| Server 58 | Model 1 | 25:19:10 | 1 | 260 | 100 | 23.25 | 28.04 | 42.32 | PML5:0.27(1.00) PML1:0.27(1.00) MG4:0.77(1.00) ANP2:0.24(1.00) ANP6:0.23(1.00) MG8:0.75(1.00) | - | - | 0.37 | 7.14 | 22.73 | 8.39 | 0.75 | 0.07 | 0.24 |
| Server 19 | Model 1 | 49:15:36 | 1 | 260 | 100 | 22.45 | 26.66 | 44.73 | PML5:0.32(1.00) PML1:0.25(1.00) MG4:0.81(1.00) ANP2:0.26(1.00) ANP6:0.23(1.00) MG8:0.81(1.00) | - | - | 0.39 | 9.23 | 24.62 | 10.31 | 0.79 | 0.11 | 0.23 |
| Princeton_TEMPLATE | Model 1 | 01:24:35 | 1 | 260 | 100 | 20.05 | 24.37 | 39.67 | PML5:0.16(1.00) PML1:0.20(1.00) MG4:0.73(1.00) ANP2:0.29(1.00) ANP6:0.28(1.00) MG8:0.71(1.00) | - | - | 0.33 | 7.24 | 26.50 | 7.35 | 0.57 | 0.07 | 0.22 |

**Figure 21.** The list of CAMEO-participating servers (only 5 of 26 are displayed) and their model scores. The rightmost column displays the model TMscore. Server60 is our contact web server.

## Conclusion and Discussion

In this paper we have presented a new deep (supervised) learning method that can greatly improve protein contact prediction. Our method distinguishes itself from previous supervised learning methods in that we employ a concatenation of two deep residual neural networks to model sequence-contact relationship, one for modeling of sequential features (i.e., sequence profile, predicted secondary structure and solvent accessibility) and the other for modeling of pairwise features (e.g., coevolution information). Ultra-deep residual network is the latest breakthrough in computer vision and has demonstrated the best performance in the computer vision challenge tasks (similar to CASP) in 2015. Our method is also unique in that we predict all contacts of a protein simultaneously, which allows us to easily model high-order residue correlation. By contrast, existing supervised learning methods predict if two residues form a contact or not independent of the other residue pairs. Our (blind) test results show that our method dramatically improves contact prediction, exceeding currently the best methods (e.g., CCMpred, Evfold, PSICOV and MetaPSICOV) by a very large margin. Even without using any force fields and fragment assembly, ab initio folding using our predicted contacts as restraints can yield 3D structural models of correct fold for many test proteins. Further, our experimental results also show that our contact-assisted models are much better than template-based models built from the training proteins of our deep model. We expect that our contact prediction methods can help reveal much more biological insights for those protein families without solved structures and close structural homologs.

Our method outperforms ECA due to a couple of reasons. First, ECA predicts contacts using information only in a single protein family, while our method learns sequence-structure relationship from thousands of protein families. Second, ECA considers only pairwise residue correlation, while our deep architecture can capture high-order residue correlation (or contact occurring patterns) very well. Our method uses a subset of protein features used by MetaPSICOV, but performs much better than MetaPSICOV mainly because we explicitly model contact patterns (or high-order correlation), which is enabled by predicting contacts of a single protein simultaneously. MetaPSICOV employs a 2-stage approach. The 1$^{st}$ stage predicts if there is a contact between a pair of residues independent of the other residues. The 2$^{nd}$ stage considers the correlation between one residue pair and its neighboring pairs, but not in a very good way. In particular, the prediction errors in the 1$^{st}$ stage of MetaPSICOV cannot be corrected by the 2$^{nd}$ stage since two stages are trained separately. By contrast, we train all 2D convolution layers simultaneously (each layer is equivalent to one stage) so that later stages can correct prediction errors in early stages. In addition, a deep network can model much higher-order correlation and thus, capture information in a much larger context.

Our deep model does not predict contact maps by simply recognizing them from PDB, as evidenced by our experimental settings and results. First, we employ a strict criterion to remove redundancy so that there are no training proteins with sequence identity >25% or BLAST E-value <0.1 with any test proteins. Second, our contact-assisted models also have better quality than homology models, so it is unlikely that our predicted contact maps are simply copied from the training proteins. Third, our deep model trained by only non-membrane proteins works very well on membrane proteins. By contrast, the homology models built from the training proteins for the membrane proteins have very low quality. Their average TMscore is no more than 0.17, which is the expected TMscore of any two randomly-chosen proteins. Finally, the blind CAMEO test indicates that our method successfully folded several targets with a new fold (e.g., 5f5pH).

We have studied the impact of different input features. First of all, the co-evolution strength produced by CCMpred is the most important input features. Without it, the top L/10 long-range prediction accuracy may drop by 0.15 for soluble proteins and more for membrane proteins. The larger performance degradation for membrane proteins is mainly because information learned from sequential features of soluble proteins is not useful for membrane proteins. The depth of our deep model is equally important, as evidenced by the fact that our deep method has much better accuracy than MetaPSICOV although we use a subset of protein features used by MetaPSICOV. Our test shows that a deep model with 9 and 30 layers have top L/10 accuracy ~0.1 and ~0.03 worse than a 60-layer model, respectively. This suggests that it is very important to model contact occurring patterns (i.e., high-order residue correlation) by a deep architecture. The pairwise contact potential and mutual information may impact the accuracy by 0.02-0.03. The secondary structure and solvent accessibility may impact the accuracy by 0.01-0.02.

An interesting finding is that although our training set contains only ~100 membrane proteins, our model works well for membrane proteins, much better than CCMpred and MetaPSICOV. Even without using any membrane proteins in our training set, our deep models have almost the same accuracy on membrane proteins as those trained with membrane proteins. This implies that the sequence-structure relationship learned by our model from non-membrane proteins can generalize well to membrane protein contact prediction. We are going to study if we can further improve contact prediction accuracy of membrane proteins by including many more membrane proteins in the training set.

We may further improve contact prediction accuracy by enlarging the training set. First, the latest PDB25 has more than 10,000 proteins, which can provide many more training proteins than what we are using now. Second, when removing redundancy between training and test proteins, we may relax the BLAST E-value cutoff to 0.001 or simply drop it. This will improve the top L/k (k=1,2,5,10) contact prediction accuracy by 1-3% and accordingly the quality of the resultant 3D models by 0.01-0.02 in terms of TMscore. We may also improve the 3D model quality by combining our predicted contacts with energy function and fragment assembly. For example, we may feed our predicted contacts to Rosetta to build 3D models. Compared to CNS, Rosetta makes use of energy function and more local structural restraints through fragment assembly and thus, shall result in much better 3D models. Finally, instead of predicting contacts, our deep learning model actually can predict inter-residue distance distribution (i.e., distance matrix), which provides finer-grained information than contact maps and thus, shall benefit 3D structure modeling more than predicted contacts.

Our model achieves pretty good performance when using around 60-70 convolutional layers. A natural question to ask is can we further improve prediction accuracy by using many more convolutional layers? In computer vision, it has been shown that a 1001-layer residual neural network can yield better accuracy for image-level classification than a 100-layer network (but no result on pixel-level labeling is reported). Currently we cannot apply more than 100 layers to our model due to insufficient memory of a GPU card (12G). We plan to overcome the memory limitation by extending our training algorithm to run on multiple GPU cards. Then we will train a model with hundreds of layers to see if we can further improve prediction accuracy or not.

# Method

## Deep learning model details

**Residual network blocks.** Our network consists of two residual neural networks, each in turn consisting of some residual blocks concatenated together. Fig. 22 shows an example of a residual block consisting of 2 convolution layers and 2 activation layers. In this figure, $X_l$ and $X_{l+1}$ are the input and output of the block, respectively. The activation layer conducts a simple nonlinear transformation of its input without using any parameters. Here we use the ReLU activation function (30) for such a transformation. Let $f(X_l)$ denote the result of $X_l$ going through the two activation layers and the two convolution layers. Then, $X_{l+1}$ is equal to $X_l + f(X_l)$. That is, $X_{l+1}$ is a combination of $X_l$ and its nonlinear transformation. Since $f(X_l)$ is equal to the difference between $X_{l+1}$ and $X_l$, $f$ is called residual function and this network called residual network. In the first residual network, $X_l$ and $X_{l+1}$ represent sequential features and have dimension $L \times n_l$ and $L \times n_{l+1}$, respectively, where L is protein sequence length and $n_l$ ($n_{l+1}$) can be interpreted as the number of features or hidden neurons at each position (i.e., residue). In the 2$^{nd}$ residual network, $X_l$ and $X_{l+1}$ represent pairwise features and have dimension $L \times L \times n_l$ and $L \times L \times n_{l+1}$, respectively, where $n_l$ ($n_{l+1}$) can be interpreted as the number of features or hidden neurons at one position (i.e., residue pair). Typically, we enforce $n_l \leq n_{l+1}$ since one position at a higher level is supposed to carry more information. When $n_l < n_{l+1}$, in calculating $X_l + f(X_l)$ we shall pad zeros to $X_l$ so that it has the same dimension as $X_{l+1}$. To speed up training, we also add a batch normalization layer (43) before each activation layer, which normalizes its input to have mean 0 and standard deviation 1. The filter size (i.e., window size) used by a 1D convolution layer is 17 while that used by a 2D convolution layer is 3×3 or 5×5. By stacking many residual blocks together, even if at each convolution layer we use a small window size, our network can model very long-range interdependency between input features and contacts as well as the long-range interdependency between two different residue pairs. We fix the depth (i.e., the number of convolution layers) of the 1D residual network to 6, but vary the depth of the 2D residual network. Our experimental results show that with ~60 hidden neurons at each position and ~60 convolution layers for the 2$^{nd}$ residual network, our model can yield pretty good performance. Note that it has been shown that for image classification a convolutional neural network with a smaller window size but many more layers usually outperforms a network with a larger window size but fewer layers. Further, a 2D convolutional neural network with a smaller window size also has a smaller number of parameters than a network with a larger window size. See https://github.com/KaimingHe/deep-residual-networks for some existing implementations of 2D residual neural network. However, they assume an input of fixed dimension, while our network needs to take variable-length proteins as input.

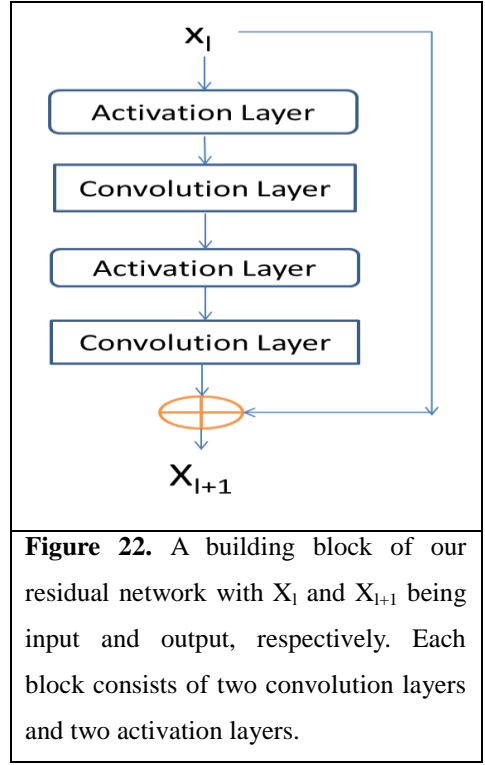

**Figure 22.** A building block of our residual network with $X_l$ and $X_{l+1}$ being input and output, respectively. Each block consists of two convolution layers and two activation layers.

Our deep learning method for contact prediction is unique in at least two aspects. First, our model employs two multi-layer residual neural networks, which have not been applied to contact prediction before. Residual neural networks can pass both linear and nonlinear information from end to end (i.e., from the initial input to the final output). Second, we do contact prediction on the whole contact map by treating it as an individual image. In contrast, previous supervised learning methods separate the prediction of one residue pair from the others. By predicting contacts of a protein simultaneously, we can easily model long-range contact correlation and high-order residue correlation and long-range correlation between a contact and input features.

**Convolutional operation.** Existing deep learning development toolkits such as Theano (http://deeplearning.net/software/theano/) and Tensorflow (https://www.tensorflow.org/) have provided an API (application programming interface) for convolutional operation so that we do not need to implement it by ourselves. See http://deeplearning.net/tutorial/lenet.html and https://www.nervanasys.com/convolutional-neural-networks/ for a good tutorial of convolutional network. Please also see (44) for a detailed account of 1D convolutional network with application to protein sequence labeling. Roughly, a 1D convolution operation is de facto matrix-vector multiplication and 2D convolution can be interpreted similarly. Let X and Y (with dimensions L×m and L×n, respectively) be the input and output of a 1D convolutional layer, respectively. Let the window size be 2w+1 and s=(2w+1)m. The convolutional operator that transforms X to Y can be represented as a 2D matrix with dimension n×s, denoted as C. C is protein length-independent and each convolutional layer may have a different C. Let $X_i$ be a submatrix of X centered at residue i (1≤ i ≤L) with dimension (2w+1)×m, and $Y_i$ be the i-th row of Y. We may calculate $Y_i$ by first flattening $X_i$ to a vector of length s and then multiplying C and the flattened $X_i$.

**Conversion of sequential features to pairwise features.** We convert the output of the first module of our model (i.e., the 1-d residual neural network) to a 2D representation using an operation similar to outer product. Simply speaking, let v={$v_1$, $v_2$, …, $v_i$, …, $v_L$} be the final output of the first module where L is protein sequence length and $v_i$ is a feature vector storing the output information for residue *i*. For a pair of residues *i* and *j*, we concatenate $v_i$ , $v_{(i+j)/2}$ and $v_j$ to a single vector and use it as one input feature of this residue pair. The input features for this pair also include mutual information, the EC information calculated by CCMpred and pairwise contact potential (45, 46).

**Loss function.** We use maximum-likelihood method to train model parameters. That is, we maximize the occurring probability of the native contacts (and non-contacts) of the training proteins. Therefore, the loss function is defined as the negative log-likelihood averaged over all the residue pairs of the training proteins. Since the ratio of contacts among all the residue pairs is very small, to make the training algorithm converge fast, we assign a larger weight to the residue pairs forming a contact. The weight is assigned such that the total weight assigned to contacts is approximately 1/8 of the number of non-contacts in the training set.

**Regularization and optimization.** To prevent overfitting, we employ $L_2$-norm regularization to reduce the parameter space. That is, we want to find a set of parameters with a small $L_2$ norm to minimize the loss function, so the final objective function to be minimized is the sum of loss function and the $L_2$ norm of the model parameters (multiplied by a regularization factor). We use a stochastic gradient descent algorithm to minimize the objective function. It takes 20-30 epochs (each epoch scans through all the training proteins exactly once) to obtain a very good solution. The whole algorithm is implemented by Theano (47) and mainly runs on GPU.

**Training and dealing with proteins of different lengths.** Our network can take as input variable-length proteins. We train our deep network in a minibatch mode, which is routinely used in deep learning. That is, at each iteration of our training algorithm, we use a minibatch of proteins to calculate gradient and update the model parameters. A minibatch may have one or several proteins. We sort all training proteins by length and group proteins of similar lengths into minibatches. Considering that most proteins have length up to 600 residues, proteins in a minibatch often have the same length. In the case that they do not, we add zero padding to shorter proteins. Our convolutional operation is protein-length independent, so two different minibatches are allowed to have different protein lengths. We have tested minibatches with only a single protein or with several proteins. Both work well. However, it is much easier to implement minibatches with only a single protein.

Since our network can take as input variable-length lengths, we do not need to cut a long protein into segments in predicting contact maps. Instead we predict contacts in the whole chain simultaneously. There is no need to use zero padding when only a single protein is predicted in a batch. Zero padding is needed only when several proteins of different lengths are predicted in a batch.

## Training and test data

Our test data includes the 150 Pfam families (5), 105 CASP11 test proteins, 76 hard CAMEO test proteins released in 2015 (Supplementary Table 1) and 398 membrane proteins (Supplementary Table 2). All test membrane proteins have length no more than 400 residues and any two membrane proteins share less than 40% sequence identity. For the CASP test proteins, we use the official domain definitions, but we do not parse a CAMEO or membrane protein into domains.

Our training set is a subset of PDB25 created in February 2015, in which any two proteins share less than 25% sequence identity. We exclude a protein from the training set if it satisfies one of the following conditions: (i) sequence length smaller than 26 or larger than 700, (ii) resolution worse than 2.5Å, (iii) has domains made up of multiple protein chains, (iv) no DSSP information, and (v) there is inconsistency between its PDB, DSSP and ASTRAL sequences (48). To remove redundancy with the test sets, we exclude any training proteins sharing >25% sequence identity or having BLAST E-value <0.1 with any test proteins. In total there are 6767 proteins in our training set, from which we have trained 7 different models. For each model, we randomly sampled ~6000 proteins from the training set to train the model and used the remaining proteins to validate the model and determine the hyper-parameters (i.e., regularization factor). The final model is the average of these 7 models.

## Protein features

We use similar but fewer protein features as MetaPSICOV. In particular, the input features include protein sequence profile (i.e., position-specific scoring matrix), predicted 3-state secondary structure and 3-state solvent accessibility, direct co-evolutionary information generated by CCMpred, mutual information and pairwise potential (45, 46). To derive these features, we need to generate MSA (multiple sequence alignment). For a training protein, we run PSI-BLAST (with E-value 0.001 and 3 iterations) to search the NR (non-redundant) protein sequence database dated in October 2012 to find its sequence homologs, and then build its MSA and sequence profile and predict other features (i.e., secondary structure and solvent accessibility). Sequence profile is represented as a 2D matrix with dimension $L \times 20$ where L is the protein length. Predicted secondary structure is represented as a 2D matrix with dimension $L \times 3$ (each entry is a predicted score or probability), so is the predicted solvent accessibility. Concatenating them together, we have a 2D matrix with dimension $L \times 26$, which is the input of our 1D residual network.

For a test protein, we generate four different MSAs by running HHblits (38) with 3 iterations and E-value set to 0.001 and 1, respectively, to search through the uniprot20 HMM library released in November 2015 and February 2016. From each individual MSA, we derive one sequence profile and employ our in-house tool RaptorX-Property (49) to predict the secondary structure and solvent accessibility accordingly. That is, for each test protein we generate 4 sets of input features and accordingly 4 different contact predictions. Then we average these 4 predictions to obtain the final contact prediction. This averaged contact prediction is about 1-2% better than that predicted from a single set of features (detailed data not shown). Although currently there are quite a few packages that can generate direct evolutionary coupling information, we only employ CCMpred to do so because it runs fast on GPU (4).

## Programs to compare and evaluation metrics

We compare our method with PSICOV (5), Evfold (6), CCMpred (4), plmDCA, Gremlin, and MetaPSICOV (9). The first 5 methods conduct pure DCA while MetaPSICOV employs supervised learning. MetaPSICOV (9) performed the best in CASP11 (31). CCMpred, plmDCA, Gremlin perform similarly, but better than PSICOV and Evfold. All the programs are run with parameters set according to their respective papers. We evaluate the accuracy of the top $L/k$ ($k$=10, 5, 2, 1) predicted contacts where L is protein sequence length. The prediction accuracy is defined as the percentage of native contacts among the top $L/k$ predicted contacts. We also divide contacts into three groups according to the sequence distance of two residues in a contact. That is, a contact is short-, medium- and long-range when its sequence distance falls into [6, 11], [12, 23], and $\geq$24, respectively.

## Calculation of Meff

Meff measures the amount of homologous information in an MSA (multiple sequence alignment). It can be interpreted as the number of non-redundant sequence homologs in an MSA when 70% sequence identity is used as cutoff. To calculate Meff, we first calculate the sequence identity between any two proteins in the MSA. Let a binary variable $S_{ij}$ denote the similarity between two protein sequences i and j. $S_{ij}$ is equal to 1 if and only if the sequence identity between i and j is at least 70%. For a protein i, we calculate the sum of $S_{ij}$ over all the proteins (including itself) in the MSA and denote it as $S_i$. Finally, we calculate Meff as the sum of $1/S_i$ over all the protein sequences in this MSA.

### 3D model construction by contact-assisted folding

We use a similar approach as described in (11) to build the 3D models of a test protein by feeding predicted contacts and secondary structure to the Crystallography & NMR System (CNS) suite (32). We predict secondary structure using our in-house tool RaptorX-Property (49) and then convert it to distance, angle and h-bond restraints using a script in the Confold package (11). For each test protein, we choose top 2L predicted contacts (L is sequence length) no matter whether they are short-, medium- or long-range and then convert them to distance restraints. That is, a pair of residues predicted to form a contact is assumed to have distance between 3.5Å and 8.0 Å. In current implementation, we do not use any force fields to help with folding. We generate twenty 3D structure models using CNS and select top 5 models by the NOE score yielded by CNS(32). The NOE score mainly reflects the degree of violation of the model against the input constraints (i.e., predicted secondary structure and contacts). The lower the NOE score, the more likely the model has a higher quality. When CCMpred- and MetaPSICOV-predicted contacts are used to build 3D models, we also use the secondary structure predicted by RaptorX-Property to warrant a fair comparison.

### Template-based modeling (TBM) of the test proteins

To generate template-based models (TBMs) for a test protein, we first run HHblits (with the UniProt20_2016 library) to generate an HMM file for the test protein, then run HHsearch with this HMM file to search for the best templates among the 6767 training proteins of our deep learning model, and finally run MODELLER to build a TBM from each of the top 5 templates.

### Author contributions

J.X. conceived the project, developed the algorithm and wrote the paper. S.W. did data preparation and analysis and helped with algorithm development and paper writing. S.S. helped with algorithm development and data analysis. R.Z. helped with data analysis. Z.L. helped with algorithm development.

### Acknowledgements

This work is supported by National Institutes of Health grant R01GM089753 to J.X. and National Science Foundation grant DBI-1564955 to J.X. The authors are also grateful to the support of Nvidia Inc. and the computational resources provided by XSEDE.

# References

1. Kim DE, DiMaio F, Yu‐Ruei Wang R, Song Y, Baker D. One contact for every twelve residues allows robust and accurate topology‐level protein structure modeling. Proteins: Structure, Function, and Bioinformatics. 2014;82(S2):208-18.

2. de Juan D, Pazos F, Valencia A. Emerging methods in protein co-evolution. Nature Reviews Genetics. 2013;14(4):249-61.

3. Weigt M, White RA, Szurmant H, Hoch JA, Hwa T. Identification of direct residue contacts in protein-protein interaction by message passing. P Natl Acad Sci USA. 2009 Jan 6;106(1):67-72.

4. Seemayer S, Gruber M, Söding J. CCMpred—fast and precise prediction of protein residue–residue contacts from correlated mutations. Bioinformatics. 2014;30(21):3128-30.

5. Jones DT, Buchan DW, Cozzetto D, Pontil M. PSICOV: precise structural contact prediction using sparse inverse covariance estimation on large multiple sequence alignments. Bioinformatics. 2012;28(2):184-90.

6. Marks DS, Colwell LJ, Sheridan R, Hopf TA, Pagnani A, Zecchina R, et al. Protein 3D structure computed from evolutionary sequence variation. PloS one. 2011;6(12):e28766.

7. Ekeberg M, Hartonen T, Aurell E. Fast pseudolikelihood maximization for direct-coupling analysis of protein structure from many homologous amino-acid sequences. J Comput Phys. 2014 Nov 1;276:341-56.

8. Kamisetty H, Ovchinnikov S, Baker D. Assessing the utility of coevolution-based residue–residue contact predictions in a sequence-and structure-rich era. Proceedings of the National Academy of Sciences. 2013;110(39):15674-9.

9. Jones DT, Singh T, Kosciolek T, Tetchner S. MetaPSICOV: combining coevolution methods for accurate prediction of contacts and long range hydrogen bonding in proteins. Bioinformatics. 2015;31(7):999-1006.

10. Ma J, Wang S, Wang Z, Xu J. Protein contact prediction by integrating joint evolutionary coupling analysis and supervised learning. Bioinformatics. 2015:btv472.

11. Adhikari B, Bhattacharya D, Cao R, Cheng J. CONFOLD: residue‐residue contact‐guided ab initio protein folding. Proteins: Structure, Function, and Bioinformatics. 2015;83(8):1436-49.

12. Wang S, Li W, Zhang R, Liu S, Xu J. CoinFold: a web server for protein contact prediction and contact-assisted protein folding. Nucleic acids research. 2016:gkw307.

13. Di Lena P, Nagata K, Baldi P. Deep architectures for protein contact map prediction. Bioinformatics. 2012;28(19):2449-57.

14. Ekeberg M, Lövkvist C, Lan Y, Weigt M, Aurell E. Improved contact prediction in proteins: using pseudolikelihoods to infer Potts models. Phys Rev E. 2013;87(1):012707.

15. Göbel U, Sander C, Schneider R, Valencia A. Correlated mutations and residue contacts in proteins. Proteins: Structure, Function, and Bioinformatics. 1994;18(4):309-17.

16. Morcos F, Pagnani A, Lunt B, Bertolino A, Marks DS, Sander C, et al. Direct-coupling analysis of residue coevolution captures native contacts across many protein families. Proceedings of the National

Academy of Sciences. 2011;108(49):E1293-E301.

17. Skwark MJ, Raimondi D, Michel M, Elofsson A. Improved contact predictions using the recognition of protein like contact patterns. PLoS Comput Biol. 2014;10(11):e1003889.

18. Wu S, Zhang Y. A comprehensive assessment of sequence-based and template-based methods for protein contact prediction. Bioinformatics. 2008;24(7):924-31.

19. Wang Z, Xu J. Predicting protein contact map using evolutionary and physical constraints by integer programming. Bioinformatics. 2013;29(13):i266-i73.

20. He K, Zhang X, Ren S, Sun J. Deep residual learning for image recognition. arXiv preprint arXiv:151203385. 2015.

21. Krizhevsky A, Hinton G. Convolutional deep belief networks on cifar-10. Unpublished manuscript. 2010;40.

22. Srivastava RK, Greff K, Schmidhuber J, editors. Training very deep networks. Advances in neural information processing systems; 2015.

23. Hinton G, Deng L, Yu D, Dahl GE, Mohamed A-r, Jaitly N, et al. Deep neural networks for acoustic modeling in speech recognition: The shared views of four research groups. IEEE Signal Processing Magazine. 2012;29(6):82-97.

24. LeCun Y, Bengio Y, Hinton G. Deep learning. Nature. 2015;521(7553):436-44.

25. Szegedy C, Liu W, Jia Y, Sermanet P, Reed S, Anguelov D, et al., editors. Going deeper with convolutions. Proceedings of the IEEE Conference on Computer Vision and Pattern Recognition; 2015.

26. Moult J, Fidelis K, Kryshtafovych A, Schwede T, Tramontano A. Critical assessment of methods of protein structure prediction (CASP)—round x. Proteins: Structure, Function, and Bioinformatics. 2014;82(S2):1-6.

27. Moult J, Fidelis K, Kryshtafovych A, Schwede T, Tramontano A. Critical assessment of methods of protein structure prediction: Progress and new directions in round XI. Proteins: Structure, Function, and Bioinformatics. 2016.

28. Haas J, Roth S, Arnold K, Kiefer F, Schmidt T, Bordoli L, et al. The Protein Model Portal—a comprehensive resource for protein structure and model information. Database. 2013;2013:bat031.

29. Pinheiro PH, Collobert R, editors. Recurrent Convolutional Neural Networks for Scene Labeling. ICML; 2014.

30. Nair V, Hinton GE, editors. Rectified linear units improve restricted boltzmann machines. Proceedings of the 27th International Conference on Machine Learning (ICML-10); 2010.

31. Monastyrskyy B, D'Andrea D, Fidelis K, Tramontano A, Kryshtafovych A. New encouraging developments in contact prediction: Assessment of the CASP11 results. Proteins: Structure, Function, and Bioinformatics. 2015.

32. Briinger AT, Adams PD, Clore GM, DeLano WL, Gros P, Grosse-Kunstleve RW, et al. Crystallography & NMR system: A new software suite for macromolecular structure determination. Acta Crystallogr D Biol Crystallogr. 1998;54(5):905-21.

33. Zhang Y, Skolnick J. Scoring function for automated assessment of protein structure template quality. Proteins: Structure, Function, and Bioinformatics. 2004;57(4):702-10.

34. Kim DE, Chivian D, Baker D. Protein structure prediction and analysis using the Robetta server. Nucleic Acids Research. 2004 Jul 1;32:W526-W31.

35. Kelley LA, Mezulis S, Yates CM, Wass MN, Sternberg MJE. The Phyre2 web portal for protein modeling, prediction and analysis. Nature protocols. 2015 Jun;10(6):845-58.

36. Källberg M, Wang H, Wang S, Peng J, Wang Z, Lu H, et al. Template-based protein structure modeling using the RaptorX web server. Nature protocols. 2012;7(8):1511-22.

37. Biasini M, Bienert S, Waterhouse A, Arnold K, Studer G, Schmidt T, et al. SWISS-MODEL: modelling protein tertiary and quaternary structure using evolutionary information. Nucleic Acids Research. 2014 Jul 1;42(W1):W252-W8.

38. Remmert M, Biegert A, Hauser A, Söding J. HHblits: lightning-fast iterative protein sequence searching by HMM-HMM alignment. Nature methods. 2012;9(2):173-5.

39. Xu JR, Zhang Y. How significant is a protein structure similarity with TM-score=0.5? Bioinformatics. 2010 Apr 1;26(7):889-95.

40. Wang S, Ma J, Peng J, Xu J. Protein structure alignment beyond spatial proximity. Sci Rep. [Research Support, N.I.H., Extramural

Research Support, U.S. Gov't, Non-P.H.S.]. 2013;3:1448.

41. Zalewski JK, Mo JH, Heber S, Heroux A, Gardner RG, Hildebrand JD, et al. Structure of the Shroom-Rho kinase complex reveals a binding interface with monomeric Shroom that regulates cell morphology and stimulates kinase activity. J Biol Chem. 2016 Oct 10.

42. Mohan S, Rizaldy R, Das D, Bauer RJ, Heroux A, Trakselis MA, et al. Structure of Shroom domain 2 reveals a three-segmented coiled-coil required for dimerization, Rock binding, and apical constriction. Mol Biol Cell. [Research Support, N.I.H., Extramural Research Support, U.S. Gov't, Non-P.H.S.]. 2012 Jun;23(11):2131-42.

43. Ioffe S, Szegedy C, editors. Batch Normalization: Accelerating Deep Network Training by Reducing Internal Covariate Shift. Proceedings of The 32nd International Conference on Machine Learning; 2015.

44. Wang S, Peng J, Ma J, Xu J. Protein secondary structure prediction using deep convolutional neural fields. Sci Rep. 2016;6.

45. Miyazawa S, Jernigan RL. Estimation of effective interresidue contact energies from protein crystal structures: quasi-chemical approximation. Macromolecules. 1985;18(3):534-52.

46. Betancourt MR, Thirumalai D. Pair potentials for protein folding: choice of reference states and sensitivity of predicted native states to variations in the interaction schemes. Protein Science. 1999;8(02):361-9.

47. Bergstra J, Breuleux O, Bastien F, Lamblin P, Pascanu R, Desjardins G, et al., editors. Theano: A CPU and GPU math compiler in Python. Proc 9th Python in Science Conf; 2010.

48. Drozdetskiy A, Cole C, Procter J, Barton GJ. JPred4: a protein secondary structure prediction server. Nucleic acids research. 2015:gkv332.

49. Wang S, Li W, Liu S, Xu J. RaptorX-Property: a web server for protein structure property

prediction. Nucleic acids research. 2016:gkw306.